\documentclass{aa}
\usepackage{txfonts} 
\usepackage{times}
\usepackage{graphicx}
\usepackage{natbib} 
\usepackage{color}
\usepackage{upgreek}
\usepackage{xspace}
\usepackage{tipa}
\usepackage{graphicx}
\usepackage{caption}
\usepackage{subcaption}

\bibpunct{(}{)}{;}{a}{}{,} % to follow the A&A style
 \makeatletter \newcommand \listoftodos{\section*{Automatic list of \textcolor{red}{Todo} items in text} \@starttoc{tdo}}
  \newcommand\l@todo[2]
    {\par\noindent \textit{#2}, \parbox{10cm}{#1}\par} \makeatother

\newcommand{\VVV}{\textcolor{black}{$\surd$}}

\title{Polycyclic aromatic hydrocarbon ionization as a tracer of gas flows through protoplanetary disk gaps}
\author{K. M. Maaskant\inst{1}\and M. Min\inst{2} \and L.B.F.M. Waters\inst{3,}\inst{2}   \and A.G.G.M. Tielens\inst{1} }%\and C. Dominik\inst{2,}\inst{5} \and  }
\institute{	Leiden Observatory, Leiden University, P.O. Box 9513, 2300 RA Leiden, The Netherlands \and 
		Anton Pannekoek Astronomical Institute, University of Amsterdam, P.O. Box 94249, 1090 GE Amsterdam, The Netherlands \and
		SRON Netherlands Institute for Space Research, Sorbonnelaan 2, 3584 CA Utrecht, The Netherlands }%\and
\abstract{Planet-forming disks of gas and dust around young stars contain polycyclic aromatic hydrocarbons (PAHs). % PAHs can exist in different charge states depending on the physical properties of the star and surrounding disk. 
}
{We aim to characterize how the charge state of PAHs can be used as a probe of flows of gas through protoplanetary gaps. In this context, our goal is to understand the PAH spectra of four transitional disks. In addition, we want to explain the observed correlation between PAH ionization (traced by the $I_{6.2}/I_{11.3}$ feature ratio) and the disk mass (traced by the 1.3 mm luminosity). 
}
{
We implement a model to calculate the charge state of PAHs in the Monte Carlo radiative transfer code MCMax. The emission spectra and ionization balance are calculated in the parameter space set by the properties of the star and the disk. 
}
{
A benchmark modeling grid is presented that shows how PAH ionization and luminosity behave as a function of star and disk properties. The PAH ionization is most sensitive to ultraviolet (UV) radiation and the electron density. In optically thick disks, where the UV field is low and the electron density is high, PAHs are predominantly neutral. Ionized PAHs trace low-density optically thin disk regions where the UV field is high and the electron density is low. Such regions are characteristic of gas flows through the gaps of transitional disks. We demonstrate that fitting the PAH spectra of four transitional disks requires a contribution of ionized PAHs in `gas flows' through the gap. 
}
{
The PAH spectra of transitional disks can be understood as superpositions of neutral and ionized PAHs. For  HD\,97048, neutral PAHs in the optically thick disk dominate the spectrum. In the cases of HD\,169142, HD\,135344\,B and Oph IRS 48, small amounts of ionized PAHs located in the `gas flows' through the gap are strong contributors to the total PAH luminosity. The observed trend in the sample of Herbig stars between the disk mass and PAH ionization may imply that lower-mass disks have larger gaps. Ionized PAHs in Ôgas flowsÕ through these gaps contribute strongly to their spectra.
}

\date{\today} % delete this line to display the current date
\begin{document}

\maketitle

%-----------------------------------------------------------------------------------------
%					INTRODUCTION
%-----------------------------------------------------------------------------------------

\section{Introduction}
%Many astrophysical objects show a rich spectrum of infrared (IR) emission features associated with Polycyclic Aromatic Hydrocarbons (PAHs, \citealt{1984LegerPuget,1985Allamandola}). PAH features are ubiquitously found in space, examples include HII regions, reflection nebulae, young stellar objects, planetary nebulae, post-asymptotic giant branch objects, nuclei of galaxies, and ultra-luminous infrared galaxies (\citealt{2008Tielens} and references therein). PAHs account for 10\%-30\% of cosmic carbon \citep{1995SnowWitt, 1989PugetLeger, 1989Allamandola}. They have been detected in cometary material during the Deep Impact mission \citep{2006Lisse} and were brought back to Earth by the Stardust mission \citep{2006Sandford}. It is therefore believed that PAHs are also important contributors to the composition of planets in habitable zones \citep{1993Kasting}. 

The infrared (IR) spectra originating from protoplanetary disks are rich and diverse. Polycyclic aromatic hydrocarbons (PAHs) are observed in disks around Herbig Ae/Be stars \citep{2001Meeus, 2004AckeAncker, 2008Keller, 2010Acke} and T Tauri stars \citep{2007bGeers, 2008Bouwman}. The strength of the PAH features decreases with stellar effective temperature. No T Tauri stars of the spectral type later than G8 shows the IR emission features in their spectra  (e.g. \citealt{2008Tielens,2011Kamp} and references therein). In addition, the strength of the IR emission features is generally much weaker relative to the total IR emission in the Herbig star spectra as compared to the diffuse ISM  \citep{2005Sloan, 2008Keller, 2010Acke, 2008Boersma}. Direct imaging observations taken in the PAH bands have shown that PAHs can be used as a tracer of the outer disk surface \citep{2006Doucet, 2006Lagage}.  Even so, the PAH emission is not dominated by PAHs on the surface of the outer disk in some transitional disks \citep{2007aGeers, 2013Maaskant}. 

After being electronically excited by ultraviolet (UV) photons, the PAHs cool by emission in the CH- and CC- stretching and bending modes. Laboratory studies and quantum chemical calculations show that many of the IR features shift in peak position, vary in width, and/or show substructure depending on ionization stage and detailed molecular structure \citep{1999HudginsAllamandola,1999Allamandola, 2009Bauschlicher, 2012Ricca}. An important parameter that influences the relative feature strength of the CH and CC modes is the effect of ionization with CC modes being carried predominantly by ions and CH modes by neutrals. Hence, the 6.2-$\upmu$m (CC--mode) to 11.2-$\upmu$m (CH--mode) ratio provides a good measure of the ionization balance of the emitting PAHs \citep{1999Allamandola}. This has been demonstrated in a variety of astrophysical environments \citep{2001Hony, 2002Peeters, 2005Rapacioli, 2008Galliano}.

\begin{figure*}[htbp]
\includegraphics[width=0.52\textwidth]{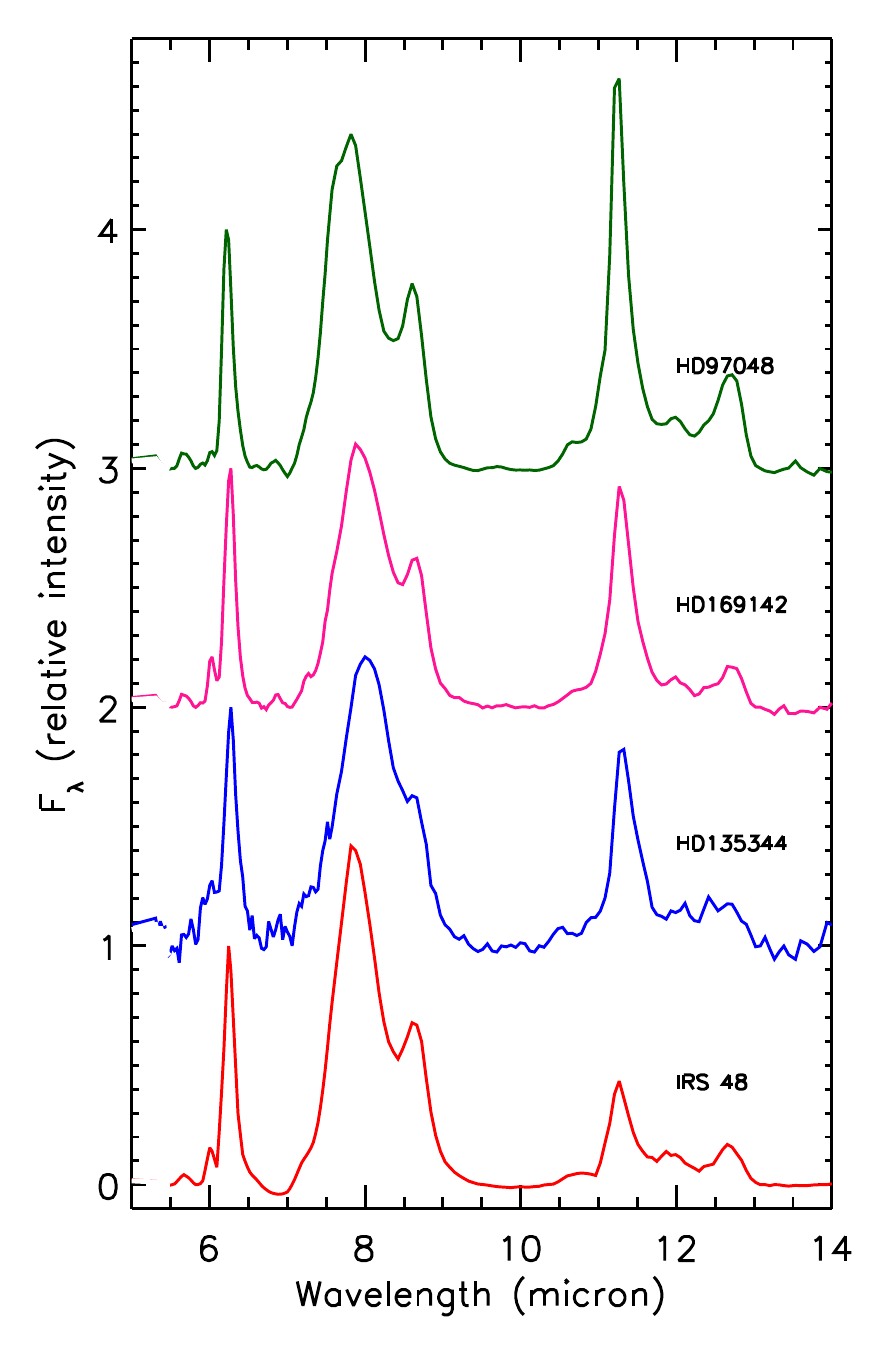}
\includegraphics[width=0.22\textwidth]{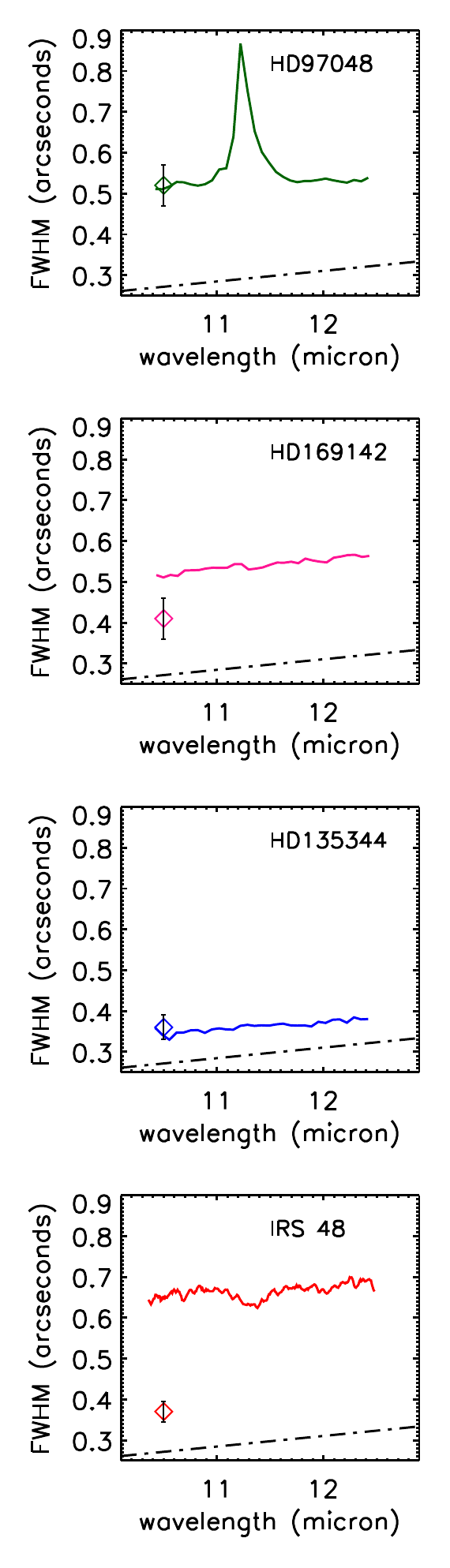}
\includegraphics[width=0.25\textwidth]{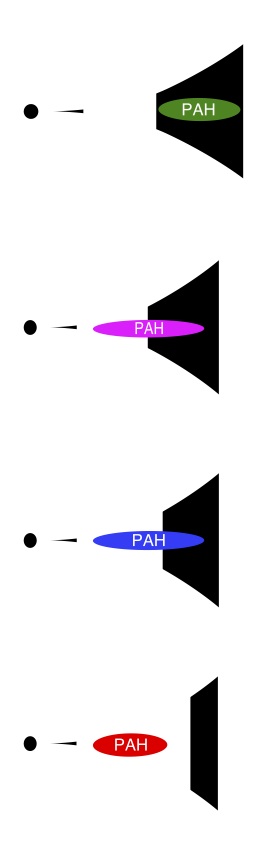}
\caption{ \label{four_objects_PAHs} Left: Spitzer/IRS spectra of the four transitional disks HD\,97048, HD\,169142, HD\,135344\,B and Oph IRS 48. The spectra are scaled relative to the 6.2-$\upmu$m PAH peak flux and show a wide range in the $I_{6.2}/I_{11.3}$ ratio; most neutral in HD\,97048 and most ionized in Oph IRS 48.  Middle: VLT/VISIR observations show that FWHM as a function of wavelength. The size of the 11.3-$\upmu$m feature is respectively larger, similar and smaller than the N-band continuum for HD\,97048, HD\,169142, HD\,135344\,B, and Oph IRS 48 (data from \citet{2013Maaskant}). The diamond symbols show the sizes of the point spread functions of the observations. The black dashed-dotted line is the diffraction limit of the telescope. Right: Sketches showing a qualitative interpretation of the spatial distribution of the dominant contribution of PAH emission, as shown by the colored ovals. Each object is given a similar color on the left, middle, and right plots. 
%A low $I_{6.2}/I_{11.3}$ ratio is indicative of neutral PAHs from the disk (e.g. HD\,97048), while a high $I_{6.2}/I_{11.3}$ ratio traces ionized PAH emission originating from the disk gap (e.g. Oph IRS 48).}
}
\end{figure*}

The intense radiation field of the central star may photo-dissociate the PAH molecules. As a result, the average PAH size increases, and the total abundance decreases \citep{2007Visser, 2012Siebenmorgen}. As a possible path for PAH-survival, \citet{2010Siebenmorgen} suggest that turbulent motions in the disk can replenish or remove PAHs from the reach of hard photons. Several modeling studies have been carried out to understand global trends in the composition and evolution of PAHs in relation to the disk \citep{2004Habart, 2007bDullemond, 2009Berne}. However, none of them are focused on understanding PAH ionization.

In this paper, we investigate the effect of disk environments on PAH charge, spectra, and luminosity. The key question is: can we use the ionization balance of PAHs as a tracer of processes in protoplanetary disks? In Section \ref{sec:obstrends}, we first identify new observational trends seen in the $I_{6.2}/I_{11.3}$ PAH feature strength ratios. Section \ref{sec:PAHmodel} describes the PAH model in the Monte Carlo radiative transfer tool MCMax. In Section \ref{sec:results}, we show how the ionization of PAHs behaves as a function of star and disk properties. We use these results to demonstrate how we can understand the PAH spectra of four transitional disks in Section \ref{sec:PAHfits}. In the discussion (Section \ref{sec:discussion}), we discuss what we can learn about disk evolution from PAH ionization. The conclusions are given in Section \ref{sec:conclusions}.

%-----------------------------------------------------------------------------------------
%					OBSERVATIONS
%-----------------------------------------------------------------------------------------

\section{New observational trends in PAH features of Herbig stars}
\label{sec:obstrends}
In this section, we discuss new observational trends of PAH ionization in relation to the disk structure of Herbig stars. For a detailed observational overview of the four transitional disks in our sample, we refer to \citet{2013Maaskant}. The PAH properties of all other Herbig stars are shown in Table \ref{table_sed_parameters} and are adopted from \citet{2010Acke}.
\subsection{Higher PAH ionization in dust depleted `gaps'}

For  HD\,97048, HD\,169142, HD\,135344\,B, and Oph IRS 48, the IRS/Spitzer spectra are shown on the left of Figure \ref{four_objects_PAHs}. The intensity of the spectra is scaled relative to the peak flux of the 6.2-$\upmu$m feature. While the general PAH profile is highly similar, a large spread is observed in the relative strength of the 11.3-$\upmu$m to the central 6.2-$\upmu$m feature. The $I_{6.2}/I_{11.3}$ ratio is lowest for HD\,97048 and increases for HD\,169142, HD\,135344\,B, and Oph IRS 48.

In contrast to HD\,97048 \citep{2006Lagage, 2006Doucet}, it was found using VLT/VISIR mid-infrared (MIR) imaging that PAH emission in Oph IRS 48 is not co-spatial to the dust continuum \citep{2007aGeers}. While the dust continuum at 18.8 $\upmu$m shows a ring structure, the PAH emission at 8.6 $\upmu$m peaks at the center. In a recent paper by \citet{2013Maaskant}, observations show that PAH emission is also not dominated by the outer flaring disk for HD\,169142 and HD\,135344\,B. These observations are shown in the middle of Figure \ref{four_objects_PAHs}, where the spatial full width at half maximum (FWHM) size of the 11.3-$\upmu$m feature can be seen as imaged by VISIR \citep{2007aGeers, 2013Maaskant}. For HD\,97048, it is clear that the 11.3-$\upmu$m feature originates in larger spatial scales than the continuum. The feature is not spatially resolved relative to the continuum in HD\,169142 and HD\,135344\,B and is smaller than the continuum in Oph IRS 48. Using radiative transfer modeling, \citet{2013Maaskant} concluded that a significant contribution must come from close to the star and possibly from within the dust depleted `gap' if the PAH feature is not larger than the continuum.  

\begin{figure}[t] 
\includegraphics[width=\columnwidth]{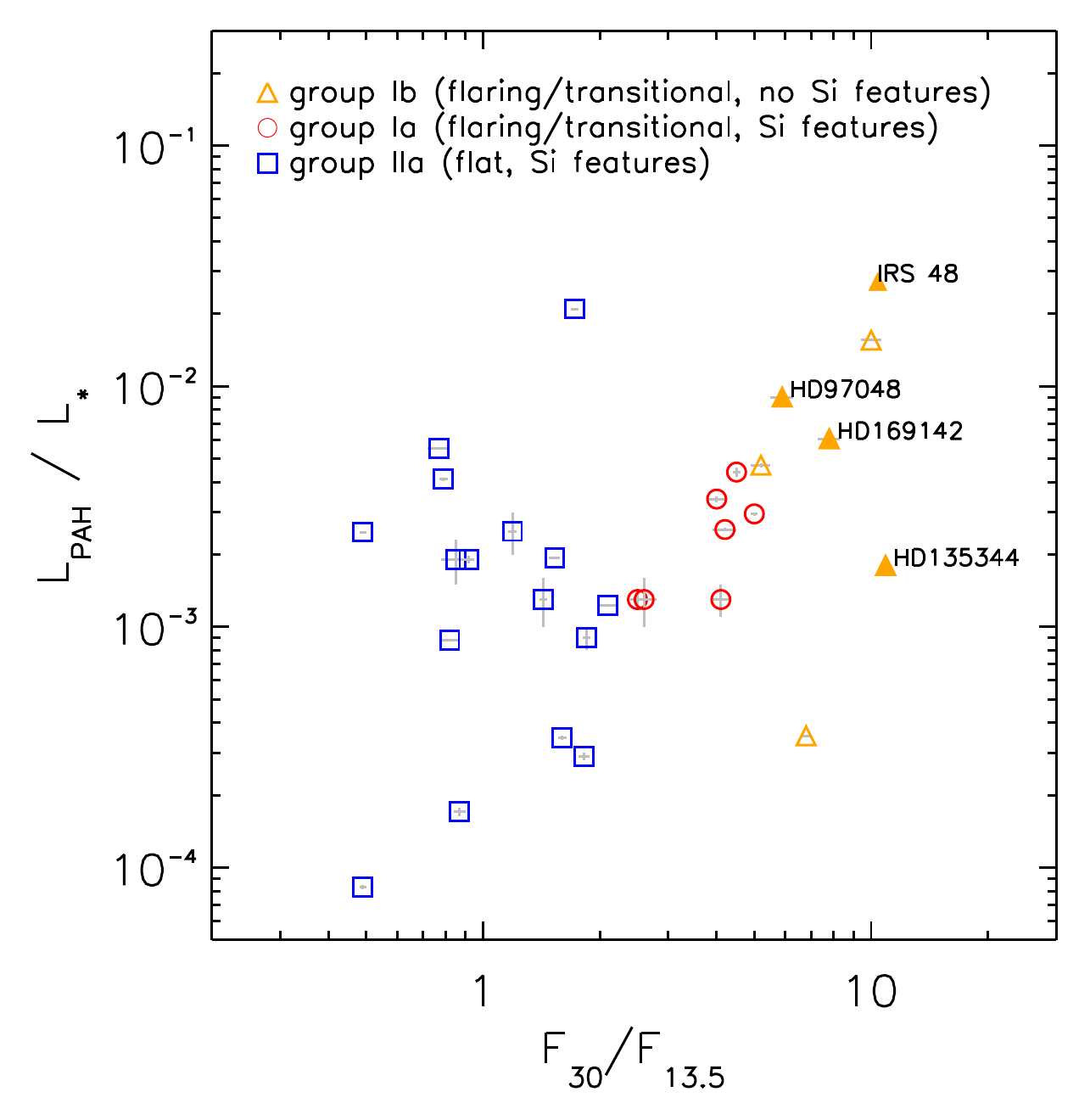}
\caption{\label{fig:MIR_vs_LPAH} The PAH luminosity ($L_{\mathrm{PAH}}/L_{*}$) of a sample of Herbig stars compared to the MIR spectral index ($F_{30}/F_{13.5}$). The MIR spectral index can be used as a tracer of disk gaps, as indicated by the symbols showing the flaring/transitional disks (group I) with silicate features (red circles), without silicate features (orange triangles), and self-shadowed disks (group II) with silicate features (blue squares). While the PAH luminosities among the objects span $\sim$2--3 orders of magnitude, the averaged luminosities of the disk groups only differ by a factor of $\sim$2--3. The parameters of all the objects are listed in Table \ref{table_sed_parameters}. } 
\end{figure}

\begin{figure}[t] 
\includegraphics[width=\columnwidth]{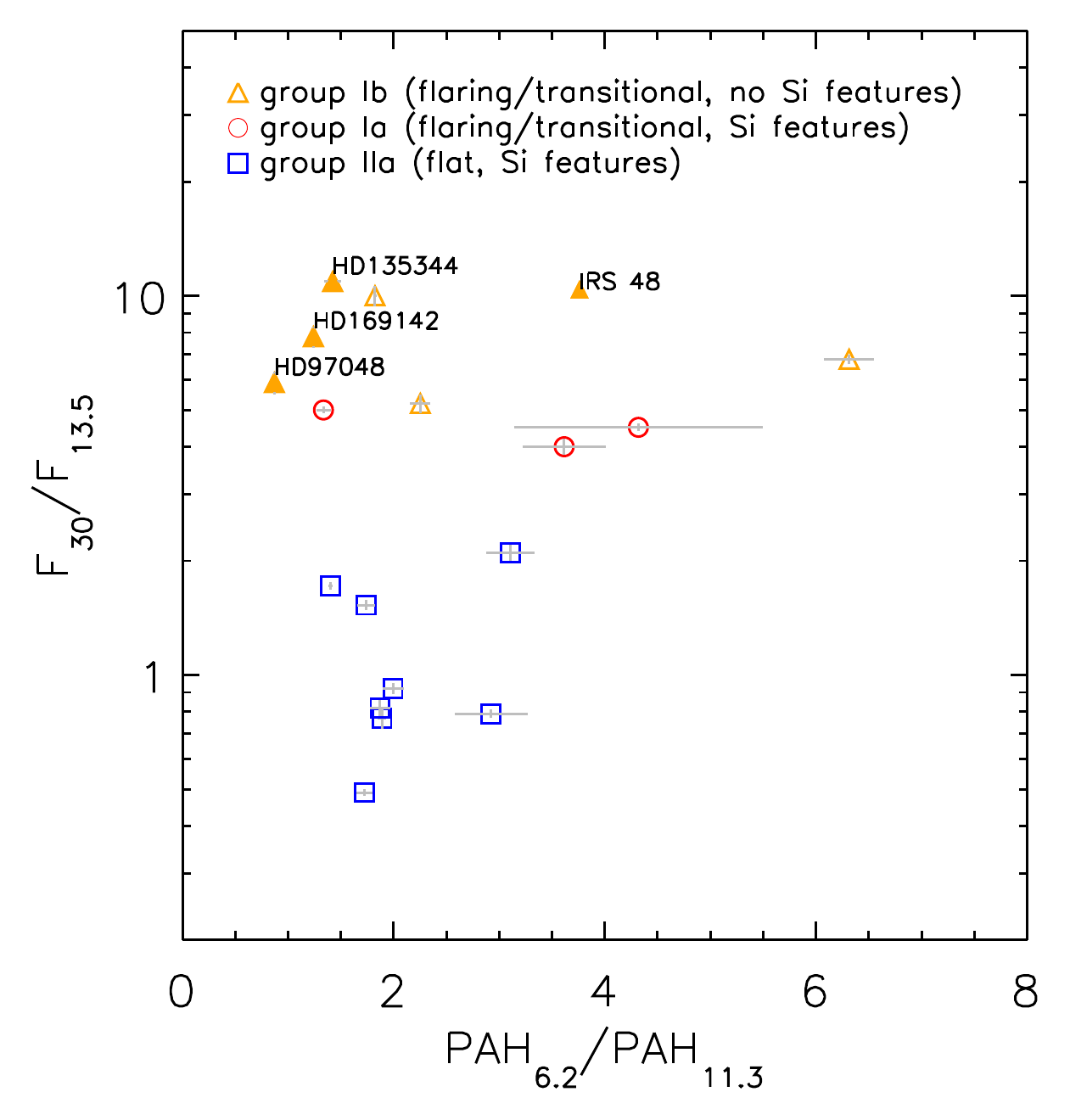}
\caption{\label{fig:ion_vs_LPAH} The $I_{6.2}/I_{11.3}$ ratio compared to the MIR spectral index ($F_{30}/F_{13.5}$). Ionized PAHs are traced by a high $I_{6.2}/I_{11.3}$ ratio. The wider range in the $I_{6.2}/I_{11.3}$ ratio in flaring/transitional disks \mbox{($\langle I_{6.2}/I_{11.3}\rangle = 2.61 \pm 1.63$)} suggests that the PAH emission originates in a wider range of physical environments while the origin of PAH emission in flat disks  \mbox{($\langle I_{6.2}/I_{11.3}\rangle   = 2.08 \pm 0.56$)} seems to be more homogeneous. The parameters of all the objects are listed in Table \ref{table_sed_parameters}. } 

\end{figure}

On the left of Figure \ref{four_objects_PAHs} sketches are shown of the disk geometries which include colored ovals to indicate the origins of the dominant PAH contributions. The sizes of the disks are taken from the models of \citet{2013Maaskant}, where the inner edge of the outer disk is set by MIR imaging. The inner disks represent the origins of near-infrared (NIR) emission in the spectral energy distributions (SEDs) though their structures are not well constrained.

The four transitional Herbig stars show a trend between the relative band strength ratio $I_{6.2}/I_{11.3}$ and the spatial origin in the disk. For HD\,97048, where the PAH emission is dominated by the outer disk, the $I_{6.2}/I_{11.3}$ ratio is low: 0.87 $\pm$ 0.01. In HD\,169142 and HD\,135344\,B, the 11.3-$\upmu$m PAH features are not extended when compared to the continuum and have $I_{6.2}/I_{11.3}$ ratios of respectively 1.24 $\pm$ 0.01 and 1.42 $\pm$ 0.08. In Oph IRS 48, the PAHS are emitted from a location that is closer to the star than the outer disk, and have a high $I_{6.2}/I_{11.3}$ ratio of  3.76 $\pm$ 0.01. Thus it seems the PAH emission which originate from the outer dusty disk has a lower $I_{6.2}/I_{11.3}$ ratio, while PAH emission from within the `gap' has a high $I_{6.2}/I_{11.3}$ ratio. Since the $I_{6.2}/I_{11.3}$ ratio can be attributed to the charge state of the PAHs \citep{1999Allamandola}, we find that the PAHs in the optically thick dust disk are predominantly neutral, while the PAHs in the disk gap are ionized.

\subsection{PAH ionization in Herbig stars}

\begin{figure}[t] 
\includegraphics[width=\columnwidth]{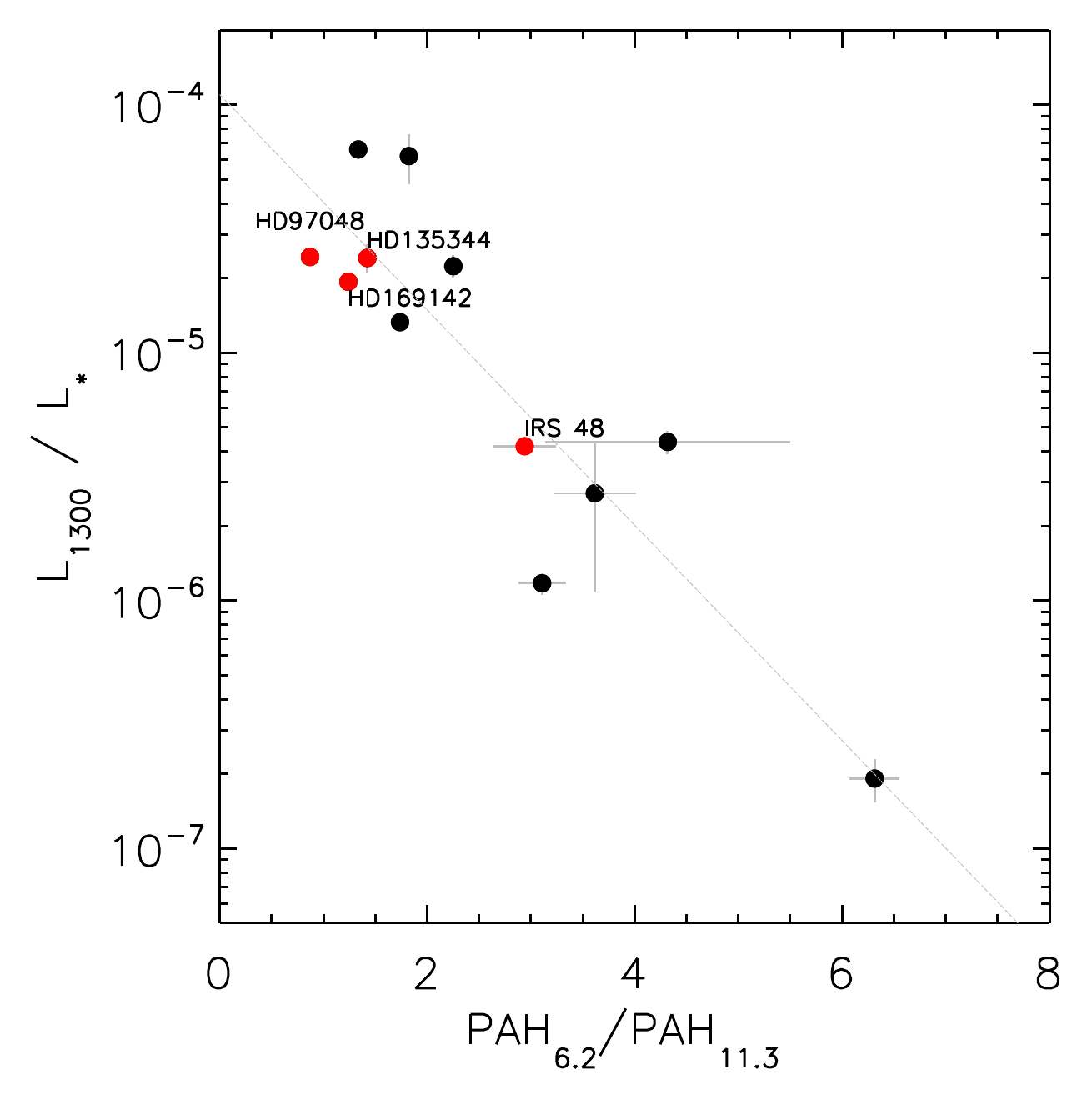}
\caption{\label{fig:6_2_divided_by_11_3_vs_L1300_divide_Lstar1300_divideL} An observational trend between the $I_{6.2}/I_{11.3}$ ratio as a function of the luminosity at 1.3 mm ($L_{1300}/L_{*}$).  The red dots show the objects studied in this paper. The gray line gives the best fit to the data (Equation \ref{eq:correlation}). The parameters of all the objects are listed in Table \ref{table_sed_parameters}. For high mm luminosities, the $I_{6.2}/I_{11.3}$ ratio is lowest (tracing neutral PAHs), while the $I_{6.2}/I_{11.3}$ ratio is highest (tracing ionized PAHs) for low mm values. }

\end{figure}

To better understand the behavior of the $I_{6.2}/I_{11.3}$ ratio for a broad range of disk geometries, we compare a large representative sample of IRS/Spitzer observations of Herbig objects (see Table \ref{table_sed_parameters}). For 47 objects, IRS/Spitzer spectra are adopted from \citet{2010Acke}. For 30 stars out of this sample, PAH luminosities were derived from detected features in the IRS/Spitzer spectrum. We discarded objects for which evidence of significant contamination from the surroundings exists in the literature. The objects are sorted by the $F_{30}/F_{13.5}$ flux ratio between 30.0-  and 13.5-$\upmu$m. This flux ratio is indicative for the disk geometry as it may trace gap-sizes in the temperature range of $\sim$200--500 K and thus allows us to search for trends as a result of the geometry of the disk. The PAH luminosities and band-strength ratio $6.2/11.3$ are included in Table \ref{table_sed_parameters}. There is no evidence that the objects in Table \ref{table_sed_parameters}, for which no PAH luminosity could be derived, represent a particular range in $F_{30}/F_{13.5}$ or $L_{1300}/L_{*}$.

For all the objects with PAH emission, Figure \ref{fig:MIR_vs_LPAH} shows the relative PAH luminosities ($L_{PAH}/L_{*}$) compared to the MIR spectral index $F_{30}/F_{13.5}$. The three different symbols indicate the SED group classifications from \citet{2001Meeus} and whether the object shows the amorphous 10- and 20-$\upmu$m silicate features in the spectrum. The blue squares (group IIa) are interpreted as the `flat', self-shadowed disks with silicate features. The red circles (group Ia) are `flaring/transitional' objects with silicate features. The orange objects (group Ib) are `flaring/transitional' objects without silicate features. The latter typically show prominent PAH spectra due to large gaps in the disk, which lower the continuum emission, thereby, increasing the contrast to the PAH features \citep{2013Maaskant}. Within the groups, $L_{\mathrm{PAH}}/L_{*}$ spans $\sim$2--3 orders of magnitude. The averaged difference between the groups is much smaller; flaring/transitional sources are, relatively speaking, only a factor of $\sim$2--3 more luminous in the fractional PAH luminosity than flat objects.

Figure \ref{fig:ion_vs_LPAH} shows the $I_{6.2}/I_{11.3}$ ratio compared to the  $F_{30}/F_{13.5}$ MIR spectral index. This figure shows that the averaged  $\langle I_{6.2}/I_{11.3}\rangle$  ratio in flaring/transitional objects \mbox{($2.61 \pm 1.63$)} is higher and has a larger spread than in flat objects \mbox{($2.08 \pm 0.56$)}. This may reflect that the PAH emission in flaring/transitional disks originates in more varied physical environments (with some cases, a strong contribution from ionized PAHs) while flat disks may be more homogeneous. In this paper we show that ionized PAH emission can arise from the gas inside the gap for transitional disks, which may arise in the gas flowing from the outer disk through the gap to the inner disk. For flat disks, there are few observational constraints on their disk geometries. However, for one particular case, there is evidence of extended PAH emission. \citet{2010Verhoeff} has shown for HD\,95881 that PAHs have a spatial extent of $\sim$100 AU and are well resolved with respect to the continuum. They, thus, confirmed the presence of an illuminated gas surface. As discussed by \citet{2007bDullemond}, the opacity in the upper layers of the disk decreases, causing a weaker far-infrared (FIR) flux  when dust decouples from the gas and settles to the mid-plane. As a consequence, PAH emission from flat disks is predicted to be stronger because the gas is still flaring. However, as Figure \ref{fig:MIR_vs_LPAH} shows, this is not what is observed, suggesting that some mechanism is not only removing small grains but also the PAHs from the disk upper layers in flat disks.

\subsection{PAH ionization and the millimeter luminosity}

For most of the sources, the 1.3 millimeter (mm) photometry is collected from the literature. An overview of all available photometry is listed in Table \ref{table_sed_parameters}. Since mm emission is optically thin and most prominently produced by cold mm-sized grains in the outer disk, its luminosity compared to the stellar luminosity can be used as a proxy of the total disk mass. Large uncertainties remain in the conversion of the sub-mm flux toward an estimate of the total disk mass, because of the unknown outer disk temperatures and the opacities of mm grains. Nevertheless, during the evolution of protoplanetary disks toward planetary systems, the disk loses most of the mass and the mm emission decreases. This makes it interesting to compare the mm fluxes of disks to other disk characteristics. 

We report a new trend seen in the PAH $I_{6.2}/I_{11.3}$ flux ratio as a function of sub-mm brightness.  Figure \ref{fig:6_2_divided_by_11_3_vs_L1300_divide_Lstar1300_divideL} shows the sub-mm luminosity at 1.3 mm ($L_{1300}/L_*$) compared to the $I_{6.2}/I_{11.3}$ PAH ratio for all Herbig stars with detected mm photometry. The $I_{6.2}/I_{11.3}$  flux ratio is higher for the sources with a lower mm flux.  The gray line gives the best fit to the data. Using least squares minimization we find the following exponential best fit to the data for the disks of Herbig stars with mm luminosities between $ 10^{-7}  \lesssim L_{1300}/L_* \lesssim 10^{-4}$

    \begin{equation}
    \label{eq:correlation}
\frac{L_{1300}}{L_*} = 1.1\times10^{-4} \exp( - \frac{I_{6.2}}{I_{11.3}}).	
\end{equation}

The PAHs in massive disks are predominantly neutral (low $I_{6.2}/I_{11.3}$  flux ratio), while PAHs in lower mass disks are ionized (high $I_{6.2}/I_{11.3}$  flux ratio).

\begin{table*}[htdp]
\tiny
\begin{center}
\begin{tabular}{lcclccrrrc}

\hline 
\hline

Target & $F_{30}/F_{13.5}$  &group & gap? & $L_{\mathrm{PAH}} / L_{*}$ 	& $I_{6.2}/ I_{11.3}$&$L_{*}$& $T_{*}$ & $d$ &	$F_{1300}$    \\ 

	 & 	$\triangledown$		&	& 	  &     &  & L$_{\odot}$ & K & pc  & mJy  \\  
\hline

HD35929 		&      0.43 $\pm$     0.01	& IIa	&\dots&						 			      \dots				&   \dots				&       131.2	&       6870	&       510		&      \dots   		 \\ 
HD72106S 	&      0.49 $\pm$     0.01	& IIa	&\dots&						   2.48  $\pm$ 0.03 $\times$ 10$^{-3}$	&1.72 $\pm$ 0.08		&       270.9	&       10750	&       277      	&     \dots 			 \\ 
HD58647 		&      0.49 $\pm$     0.01	& IIa	&\dots&						  8.30 $\pm$ 0.20 $\times$ 10$^{-5}$     	&  \dots				&       270.9	&       10750	&       277      	&      \dots   		 \\ 
HD190073 	&      0.75 $\pm$     0.02	& IIa	&\dots&						       \dots							& \dots				&       61.9		&       8990	&       290      	&      \dots   		 \\ 
HD98922 		&      0.75 $\pm$     0.03	& IIa	&\dots&						       \dots							&  \dots				&       855.7	&       10500	&       538      	&      \dots   		 \\ 
Wray1484 	&      0.76 $\pm$     0.02	& Ia	&\dots&						       \dots							&  \dots				&       1339.9	&       30000	&       750      	&      \dots   		\\ 
HD95881 		&      0.77 $\pm$     0.05	& IIa	&\dots&						   5.52 $\pm$ 0.04 $\times$ 10$^{-3}$ 	&   1.88 $\pm$ 0.04		&       7.7		&       8990	&       118      	&      \dots   		 \\ 
NX Pup 		&      0.78 $\pm$     0.02	& IIa	&\dots&						       \dots							&  \dots				&       19.6		&       7290	&       450      	&   	  \dots 		\\ 
HD50138 		&      0.78 $\pm$     0.04	& IIa	&\dots&						       \dots							& \dots				&       423.5	&       12230	&       289      	&      \dots   		 \\ 
VV Ser 		&      0.79 $\pm$     0.02	& IIa	&\dots&						   4.11 $\pm$ 0.06 $\times$ 10$^{-3}$ 	&   2.92 $\pm$ 0.35		&       19.1		&       9000	&       330      	&     \dots 			 \\ 
HD85567 		&      0.82 $\pm$     0.04	& IIa	&\dots&						   8.80 $\pm$ 0.10 $\times$ 10$^{-4}$ 	&      1.87 $\pm$  0.10 	&       328.3	&       12450	&       480      	&      \dots   		 \\ 
HD37806 		&      0.85 $\pm$     0.07	& IIa	&\dots&						     1.90 $\pm$ 0.40 $\times$ 10$^{-3}$	&   \dots				&       145.3	&       9480	&       470      	&      \dots   		 \\ 
BF Ori 		&      0.87 $\pm$     0.03	& IIa	&\dots&						   1.71 $\pm$ 0.07 $\times$ 10$^{-4}$ 	&  \dots				&       34.1		&       8985	&       510      	&   \dots			 \\ 
HD101412 	&      0.92 $\pm$     0.03	& IIa	&\dots&						   1.90 $\pm$ 0.07 $\times$ 10$^{-3}$ 	&  1.99 $\pm$ 0.10		&       5.1		&       9960	&       118      	&      \dots   		 \\ 
HD144668 	&      0.96 $\pm$     0.02	& IIa	&\dots& 						      \dots							&    \dots				&       87.5		&       7930	&       208      	&      20 $\pm$ 16  $^{o}$ 		 \\ 
KK Oph 		&       1.04 $\pm$     0.02	& IIa	&\dots&						       \dots							&    \dots				&      0.74		&       8030	&       145      	&   52 $\pm$  15 $^{o}$	 \\ 
HD37258 		&       1.10 $\pm$     0.03	& IIa	&\dots&						       \dots							&  \dots				&       36.2		&       8970	&       510      	&      \dots   		 \\ 
HD31648 		&       1.19 $\pm$     0.03	& IIa	&\dots&   2.50 $\pm$ 0.50 $\times$ 10$^{-3}$ 							&    \dots				&       13.7		&       8720	&       131      	&   250 $\pm$ 15 $^{r,\dagger}$  \\ 
HD104237 	&       1.28 $\pm$     0.03	& IIa	&\dots&						       \dots							&  \dots				&       34.7		&       8405	&       116      	&   92 $\pm$ 19 $^{o}$ 	\\ 
HD150193 	&       1.42 $\pm$     0.05	& IIa	&\dots&						       \dots							&  \dots				&       24.0		&       8990	&       150      	&   45 $\pm$ 12 $^{r,\dagger}$	  \\ 
HD244604 	&       1.43 $\pm$     0.03	& IIa	&\dots&						   1.30 $\pm$ 0.30 $\times$ 10$^{-3}$ 	&    \dots				&       23.6		&       8730	&       340      	&      \dots   		 \\ 
HD142666 	&       1.53 $\pm$     0.05	& IIa	&\VVV$^{ \hspace{1mm}a}$&   1.94 $\pm$ 0.30 $\times$ 10$^{-3}$ 			&    1.74 $\pm$ 0.09		&       14.4		&       7580	&       145      	&   127 $\pm$ 9 $^{m}$ 	  \\ 
WW Vul 		&       1.60 $\pm$     0.04	& IIa	&\dots&						   3.47 $\pm$ 0.07 $\times$ 10$^{-4}$ 	&  \dots				&       15.8		&       8430	&       440      	&\dots			 \\ 
RR Tau 		&       1.72 $\pm$     0.04	& IIa	&\dots&						      2.10 $\pm$ 0.01 $\times$ 10$^{-2}$ 	& 1.40 $\pm$ 0.02		&       2.2		&       8460	&       160      	&  \dots 			\\ 
HD144432 	&       1.82 $\pm$     0.06	& IIa	&\VVV$^{ \hspace{1mm} b}$&   2.92 $\pm$ 0.10 $\times$ 10$^{-4}$ 		&    \dots				&       10.1		&       7345	&       145      	&  44 $\pm$   10 $^p$	 \\ 
HD37357 		&       1.85 $\pm$     0.04	& IIa	&\dots&						   9.00 $\pm$ 1.00 $\times$ 10$^{-4}$ 	& \dots				&       85.1		&       9230	&       510      	&      \dots 			\\ 
HD163296 	&       2.00 $\pm$      0.10	& IIa	&\dots&						       \dots	 						&    \dots				&       23.3		&       8720	&       122      	&  743 $\pm$   15 $^{o}$	 \\ 
HD145263 	&       2.00 $\pm$      0.10	& IIa	&\dots&						       \dots	 						&   \dots				&       3.8		&       7200	&       116      	&  \dots 			 \\ 
HD35187 		&       2.10 $\pm$      0.10	& IIa	&\dots&						   1.23 $\pm$ 0.01 $\times$ 10$^{-3}$ 	& 3.11$\pm$ 0.23		&       27.4		&       8970	&       150      	&   20 $\pm$ 2 	$^{t}$ \\ 
HD203024 	&       2.30 $\pm$      0.20	& Ia	&\dots&						       \dots							&  \dots				&       99.2		&       8200	&       620      	&      \dots   		 \\ 
HD179218 	&       2.40 $\pm$      0.20	& Ia	&\dots&						       \dots	  						&   \dots				&       79.2		&       9810	&       244      	&   71 $\pm$   7	 $^{s}$ \\ 
HD250550 	&       2.50 $\pm$      0.10	& Ia	&\dots&					   1.30 $\pm$ 0.10 $\times$ 10$^{-3}$ 		&  \dots				&       20.9		&       10700	&       280      	&     \dots 			 \\ 
HD259431 	&       2.55 $\pm$     0.05	& Ia	&\dots&						       \dots							&   \dots				&       8294.5	&       25400	&       800      	&      \dots   		 \\ 
HD38120 		&       2.60 $\pm$      0.20	& Ia	&\dots& 				  1.30 $\pm$ 0.30 $\times$ 10$^{-3}$ 			& \dots				&       89.3		&       10500	&       510      	&      \dots   		 \\ 
HD100546 	&       3.50 $\pm$      0.20	& Ia	&\VVV$^{ \hspace{1mm} c}$&       \dots								&  \dots				&       32.9		&       10500	&       103      	&   465 $\pm$ 20 $^{o}$ 	 \\ 
HD37411 		&       4.00 $\pm$      0.20	& Ia	&\dots&				     3.40 $\pm$ 0.10 $\times$ 10$^{-3}$			&   3.62 $\pm$ 0.39		&       34.4		&       9100	&       510      	&    5 $\pm$ 3 $^{o}$ 	 \\ 
HD36112 		&       4.10 $\pm$      0.20	& Ia	&\VVV$^{ \hspace{1mm} d}$&  1.30 $\pm$ 0.20 $\times$ 10$^{-3}$ 			&  \dots				&       22.2		&       7850	&       205      	&   72 $\pm$   13 $^{s}$ 	 \\ 
HD139614 	&       4.20 $\pm$      0.30	& Ia	&\VVV$^{ \hspace{1mm} e}$&   2.54 $\pm$ 0.04 $\times$ 10$^{-3}$ 		&\dots				&       8.6		&       7850	&       140      	&   242 $\pm$ 15 $^{ m}$ 	 \\ 
AB Aur 		&       4.50 $\pm$      0.10	& Ia	&\VVV$^{ \hspace{1mm} f}$&   4.40 $\pm$ 0.20 $\times$ 10$^{-3}$ 			&   4.32 $\pm$ 1.18		&       46.2		&       9520	&       144      	&   136  $\pm$ 15 $^{ r}$	 \\ 
HD142527 	&       5.00 $\pm$      0.10	& Ia	&\VVV$^{ \hspace{1mm}g}$&   2.95 $\pm$ 0.05 $\times$ 10$^{-3}$ 			&   1.34 $\pm$ 0.07  		&       50.6 		&       6260	&       198      	&   1190 $\pm$ 33 $^{p}$ 	 \\ 
T CrA 		&       5.00 $\pm$      0.30	& Ia	&\dots& 						     	\dots							&  \dots				&      0.4		&       7200	&       130      	&     219 $\pm$ 11 $^{o}$   \\ 
HD100453 	&       5.20 $\pm$      0.30	& Ib	&\VVV$^{ \hspace{1mm} h}$&   4.70 $\pm$ 0.10 $\times$ 10$^{-3}$ 		&   2.25 $\pm$ 0.10		&       8.0		&       7390	&       112      	&    200 $\pm$ 21  $^{q,\dagger}$  \\ 
HD97048 		&       5.90 $\pm$      0.40	& Ib	&\VVV$^{ \hspace{1mm} i}$&   9.01 $\pm$ 0.03 $\times$ 10$^{-3}$ 			&    0.87 $\pm$ 0.01		&       40.7		&       10010	&       158      	&  452 $\pm$ 34 $^{o}$ 	 \\ 
HD141569 	&       6.80 $\pm$      0.20	& Ib	&\VVV$^{ \hspace{1mm} j}$&   3.53 $\pm$ 0.02 $\times$ 10$^{-4}$ 			&  6.31 $\pm$ 0.24		&       18.3		&       9520	&       99		&   5 $\pm$ 1 $^p$		 \\ 
HD169142 	&       7.80 $\pm$      0.50	& Ib	&\VVV$^{ \hspace{1mm} k}$&   6.05 $\pm$ 0.02 $\times$ 10$^{-3}$ 		&     1.24 $\pm$ 0.01		&       15.3		&       8200	&       145      	&   197 $\pm$   15 $^{m}$	 \\ 
HD34282 		&       10.00 $\pm$      0.60	& Ib	&\VVV$^{ \hspace{1mm} h}$&    1.55 $\pm$ 0.01 $\times$ 10$^{-2}$ 		&1.82 $\pm$ 0.02		&       3.1		&       8720	&       164      	&      100 $\pm$ 23 $^{t}$  \\ 
Oph IRS 48	&       10.40 $\pm$      0.30	& Ib	&\VVV$^{ \hspace{1mm} l}$&    2.64 $\pm$ 0.02 	 $\times$ 10$^{-2}$		&    2.94 $\pm$ 0.01		&       14.3		&       10000	&       120      	&  60    $\pm$ 10 $^{n}$ 	 \\  
HD135344B 	&       10.90 $\pm$      0.30	& Ib	&\VVV$^{ \hspace{1mm} i}$&   1.80 $\pm$ 0.02 $\times$ 10$^{-3}$ 			& 1.42 $\pm$ 0.08		&       8.3		&       6590	&       140      	&   142 $\pm$ 19 $^{ m}$  \\

\hline

\end{tabular}
\end{center}
\caption{\label{table_sed_parameters} Stellar and disk parameters of the Herbig stars in our sample. The targets are sorted by the $F_{30}/F_{13.5}$ MIR spectral index which naturally splits most of the flat disks (group IIa) from the flaring/transitional disks (group Ia and Ib). The PAH characteristics and the stellar parameters are adopted from \citet{2010Acke}. Complementary mm photometry is taken from the literature. The $\dagger$ indicates that the 1.3 mm flux is extrapolated from other (sub-)mm values using the spectral slope between the FIR and the (sub-)mm fluxes. References: \textbf{a)} \citealt{2013Schegerer}. \textbf{b)} \citealt{2012Chen}. \textbf{c)} \citealt{2003Bouwman}. \textbf{d)} \citealt{2010Isella}. \textbf{e)} \citealt{2013Matter}. \textbf{f)} \citealt{2005Pietu}. \textbf{g)} \citealt{2006Fukagawa}. \textbf{h)} Khalafinejad et al. in prep. \textbf{i)} \citealt{2013Maaskant}. \textbf{j)} \citealt{1999Augereau}. \textbf{k)} \citealt{2012Honda}. \textbf{l)} \citealt{2007aGeers}. \textbf{m)} \citealt{1996Sylvester}. \textbf{n)} \citealt{2007Andrews}. \textbf{o)} \citealt{1994Henning}.  \textbf{o)} \citealt{2001Sylvester}. \textbf{p)} \citealt{1995Walker}. \textbf{q)} \citealt{2002meeus}. \textbf{r} \citealt{2011Sandell}. \textbf{s)} \citealt{2000Mannings}. \textbf{t)} \citealt{2004Natta}. }
\end{table*}

%-----------------------------------------------------------------------------------------
%					PAHs in MCMAX
%-----------------------------------------------------------------------------------------

\section{PAHs in the radiative transfer code MCMax}
\label{sec:PAHmodel}

In this section we present our PAH ionization model and discuss how this is implemented in the radiative transfer code MCMax.

\subsection{General description of PAH model} 
We model the disks using the radiative transfer code MCMax \citep[see][]{2009Min}. MCMax is a code that not only solves for the radiative transfer but also solves the disk structure. For the continuum radiative transfer, it uses the Monte Carlo radiative transfer scheme of \citet{2001Bjorkman} with direct re-emission and temperature correction. This scheme cannot be directly applied to the PAH molecules, since they are not in thermal equilibrium. Therefore, we need to compute the emission spectra of the thermal equilibrium grains first and iterate the radiative transfer to include the radiative transfer effect on the emissivities of the PAHs.

\begin{figure}[t]
\includegraphics[width=\columnwidth]{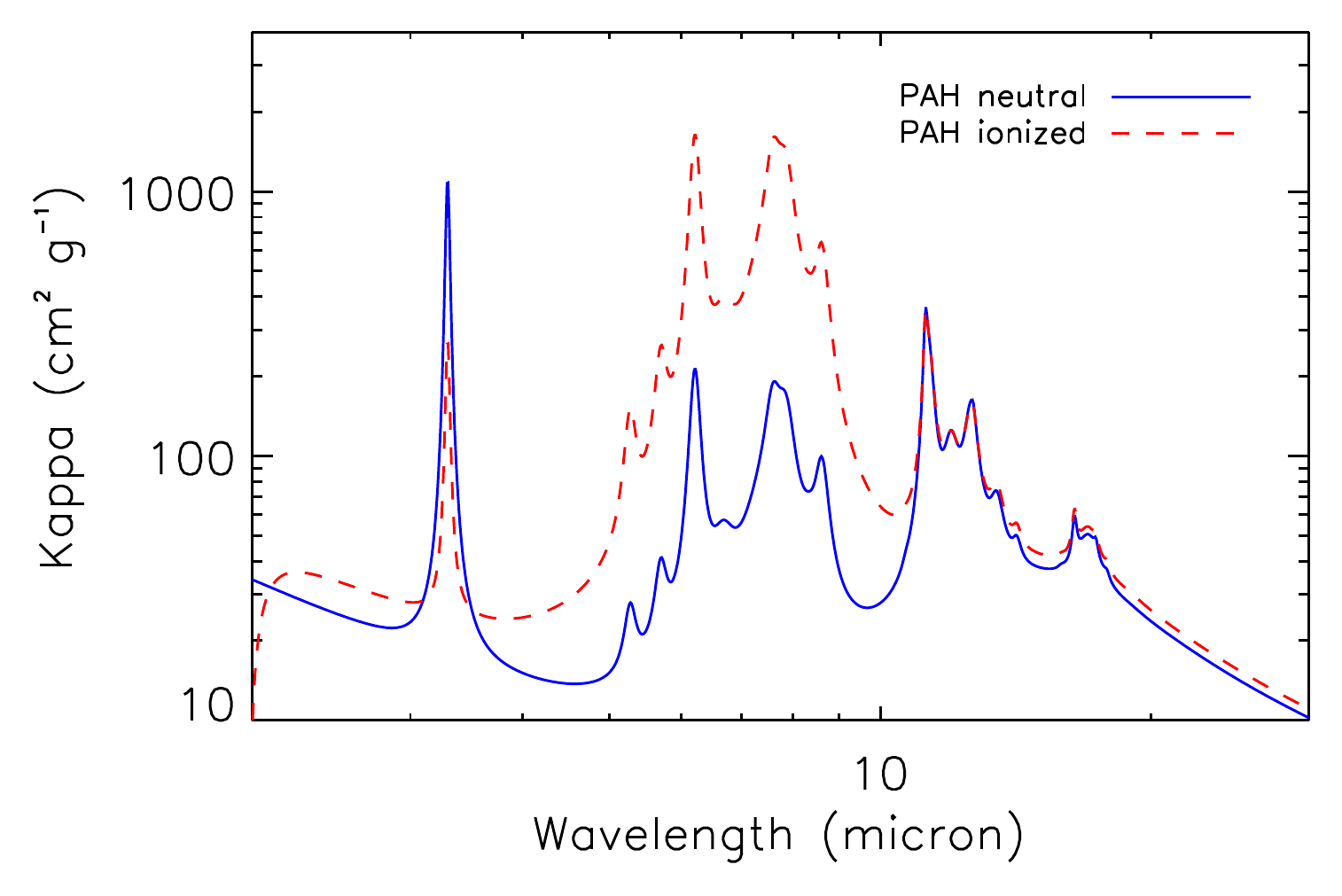}
\caption{\label{fig:kappas} Opacities for neutral PAHs (solid blue line) and ionized PAHs (dashed red line) with $N_{C} = 100$. Computed using the method described by \citet{2007DraineLi}    }
\end{figure}

The emission spectra of large molecules, like PAHs or very small dust grains, can be computed using the continuous thermal approximation \citep[][]{1989GuhathakurtaDraine, 2001DraineLi}. In this approximation it is assumed that the emission is well described by a single temperature  at any time. This temperature fluctuates strongly due to transient heating events and fast cooling. A PAH molecule cools down significantly before it is hit by the next energetic photon, which heats it again. These fluctuation, and the corresponding temperature distribution can be computed from the statistics of the radiation field and the heating and cooling properties of the particles. This method fully accounts for multi-photon heating events, or, the case where the particle is not yet cooled down completely before it is hit again. This is important for larger particles with higher heat capacities (i.e., longer cooling times) and for the intense radiation field close to the star. For the temperature distribution of the PAHs at each location in the disk is solved using a Monte Carlo method, simulating the transient heating events. An analytical method, using the solution of a set of linear equations to solve for the temperature distribution, is also implemented in MCMax \citep[also used in][]{2007bDullemond}. The two methods perfectly agree on the emerging PAH spectrum.

Once the spectral shape of the emissivity of the PAHs is computed using the temperature distribution, the PAHs can be included in the Monte Carlo radiative transfer in a very similar way to the other dust components. For ach time a photon package in the Monte Carlo radiative transfer run is absorbed by a PAH molecule, we make sure that it is re-emitted with the same energy and with the spectral signature computed using the method above. In this way, we still strictly conserve energy, ensuring a quick convergence of the method.

The optical properties of the PAHs are computed using the method by \citet{2001DraineLi, 2007DraineLi}. The adopted opacities for neutral and ionized PAHs are shown in Figure \ref{fig:kappas} (with $N_{C} = 100$, the number of carbon atoms in the PAH molecule, and PAH radius $a_{PAH}=6$ {\AA}). The PAH optical properties are constructed from a number of laboratory and observational studies. Thus our PAH model represents the emission from various PAH molecules and is designed to conform to observed PAH properties in the ISM. We use a single sized ($N_{C}$=100) PAH molecule with two possible charge states. This PAH model is sufficient for the goal of our paper, which is to provide a tool to study the effect of the star and disk properties on the charge state of PAH molecules and its influence on the resulting IR spectra. PAH processing may also play a role in protoplanetary disks, either at the molecular level or in a disk environment. Specifically, the PAH destruction is predicted to act on the smallest PAH molecules \citep{2007Visser}, but the PAHs can be replenished from deeper layers of the disk by turbulence \citep{2012Siebenmorgen}.

The $I_{6.2}/I_{11.3}$ $\upmu$m band ratio may also be affected by dehydrogenation of the PAH molecule. The effect of ionization versus dehydrogenation on the $I_{6.2}/I_{11.3}$ ratio are reviewed in \citet{2008Tielens}. The hydrogen coverage on a PAH is a balance between dehydrogenation of highly excited PAHs following UV absorption and hydrogenation reactions with atomic hydrogen \citep{1987Tielens, 2003LePage}. Theoretically and experimentally, it has been well established that dehydrogenation is an `on/off' switch; for example, when dehydrogenation becomes important, a PAH quickly loses all its hydrogens; over a small range of $G_0 / n_H$  \citep{1997Ekern, 2005Tielens, 2012BerneTielens, 2013Zhen}. Hence, for any PAH, there is a critical $G_0 / n_H$ ratio and a PAH becomes fully hydrogenated (completely dehydrogenated) if the $G_0 / n_H$ ratio is less (larger) than this ratio. As the unimolecular dissociation rate depends exponentially on the internal excitation temperature and, therefore, in a given radiation field, on the size of the PAH, PAHs with a size smaller than a critical size are fully stripped of all their H (and do not contribute emission in any C-H mode), while PAHs larger than this size are fully hydrogenated. Small fully dehydrogenated PAHs are quickly destroyed through C$_2$ loss (or isomerized to cages and fullerenes) by further UV photon absorption  \citep{2003Joblin, 2013Zhen} and this is expected to be the dominant photo destruction route for small PAHs \citep{2012BerneTielens, 2013Montillaud}. Hence, as small dehydrogenated PAHs are quickly destroyed, the $I_{6.2}/I_{11.3}$ $\upmu$m band ratio would show little variation. Moreover, if dehydrogenation were important, one would expect to see the solo/duo/trio ratio change in the direction of a larger solo/trio ratio \citep{1993Schutte}, which is not what is observed \citep{2010Acke, 2001Hony}. For these reasons, the $I_{6.2}/I_{11.3}$ ratio have been widely accepted as variations in the ionization state of the emitting PAHs, and we follow this analysis here.

The PAH-to-dust mass fraction is difficult to constrain as PAHs may take part in various processes. The PAHs may be prone to photo-destruction (e.g. \citealt{1989GuhathakurtaDraine, 2003LePage, 2007Visser}). Moreover, it has been argued that PAHs may take part in the dust coagulation process \citep{2007bDullemond} or replenished by mixing processes in the disk \citep{2010Siebenmorgen}. Observational evidence of the role of PAHs in these processes are not conclusive and it is presently unclear how important these routes are.  As for PAH destruction, very small, catacondensed PAHs can also lose  C$_2$H$_2$ through UV photon absorption while the H-loss channel dominates. However, the dominant photochemical destruction channel for the carbon skeleton for large pericondensed PAHs is through complete dehydrogenation followed by C$_2$ loss \citep{1997Ekern,2003Joblin,2013Zhen}. Larger PAHs are notoriously difficult to destroy \citep{1994Jochims}. Here, we treat the PAH/dust abundance ratio as a free parameter to be determined through our model fits of the observed SEDs.

%As for PAH destruction, very small PAHs can rapidly lose C$_2$H$_2$ through UV photon absorption. It is however not likely that this is the dominant photochemical destruction channel for the carbon skeleton because the critical size at which the carbon loss sets in is much smaller than for hydrogen loss. After extreme dehydrogenation, the resulting carbon chains and rings may grow again via carbon insertion. However, these species are then similarly susceptible to C$_2$H$_2$ loss, a process that may maintain the PAH size below a limit of $N_C \sim30$ \citep{2003LePage}. Larger PAHs are notoriously difficult to destroy \citep{1994Jochims}. } 

\subsection{The ionization balance}
The charge of PAHs is determined by the balance between photo-ionization and electron recombination. The description of the ionization balance of PAHs is largely adopted from \citet{1994BakesTielens} and \citet{2005Tielens}. Here, we provide a brief summary. The ionization balance of PAHs is controlled by the parameter $\gamma$:
    \begin{equation}
\gamma=G_0T^{1/2}/n_e 
\end{equation}
with $G_0$  as the intensity of the radiation field between 6.0 and 13.5 eV in units of the Habing field $u^{Hab}_{UV}$, $T$ the gas temperature, and $n_e$ the electron density.   

\begin{figure}[t]
\includegraphics[width=0.9\columnwidth]{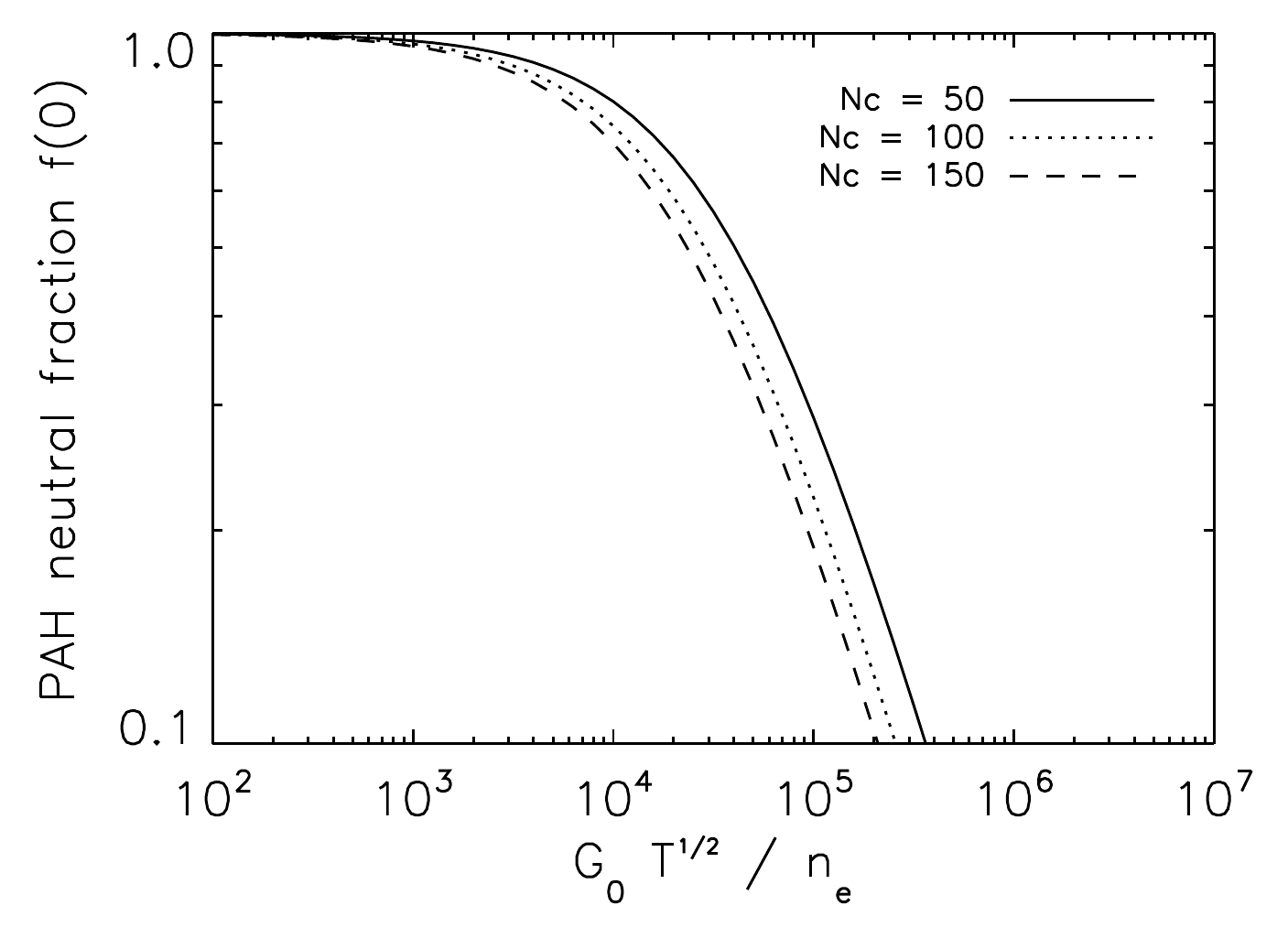}
\caption{\label{fig:f0_gamma}  The fraction of neutral PAH molecules $f(0)$ is given by the UV field, the temperature, and the electron number density and is controlled by the ionization parameter $G_0 T^{1/2}/ n_e$. In our models, we assume a PAH size of $N_{C}=100$.  }
\end{figure}

The intensity of the UV radiation field inside the disk can be characterized by $G_0= u_{UV} / u^{Hab}_{UV}$, where 
    \begin{equation}
u_{UV} =  \int_{6 eV}^{13.5  eV} u_{\nu} d\nu =  \int_{6 eV}^{13.5  eV}(h/c)h\nu N_{ph} d\nu.	
\end{equation}
Here $u^{Hab}_{UV} = 5.33 \times 10^{-14}$ erg cm$^{-3}$ is the energy density in the UV mean interstellar radiation field \citep{1968Habing}, and $N_{ph}$  is the number of photons in units of cm$^{-2}$ s$^{-1}$ erg$^{-1}$. The contribution from the interstellar radiation field is also taken into account. Regions in the disk, where the optical depth is high may not receive enough photons between 6 and 13.6 eV to calculate $G_0$. In those cases, $G_0$ is set to a lower limit value of 10$^{-6}$. Because of their high temperatures, Herbig stars generate strong photospheric UV fluxes. Additional UV excess due to accretion is relatively unimportant for these stars because of the low accretion rates in most targets. The effective temperature can, therefore, be considered as a direct indicator of the stellar UV radiation field.

The surface layer of the disk is essentially a photodissociated region \citep{2009Woitke, 2010Kamp} and the electron abundance, $x_e = n_e/n_H$, can be set equal to the C$^+$ abundance, based on an equilibrium between the photoionization of neutral C and the recombination of C$^+$ \citep{1985Tielens}. We adopt an ISM abundance of $n_C= 1.5\times$10$^{-4}n_0$ \citep{1996Cardelli} with $n_0$ the gas density in cm$^{-3}$. The C\,/\,C$^+$ ratio changes with depth but carbon is largely ionized whenever there are UV photons around\citep{2005Tielens}. Therefore, we assume that ionized carbon with a fixed abundance is the dominant supplier of free electrons in the region where PAH emission originates,. 

At higher altitudes in the disk surface, the gas temperature may exceed the dust temperature (e.g., \citealt{2004Kamp, 2005NomuraMillar}). However, it is beyond the scope of this paper to fully solve the thermo-chemistry balance. As the temperature only enters through a square root, we adopt $T_{gas}=T_{dust}$ in our model studies.

%through scattering and multi-photon events, UV and optical photons excite the PAH molecules well under the radial $\tau_{UV}=1$ surface deeper in the disk. In this region, gas heating via UV radiation is less important, and the densities are higher, thus the gas and dust are more likely to have a similar temperature. Therefore we can assume that PAH emission is dominated by disk layers where $T_{gas}=T_{dust}$.

%PAHs may have an important effect on the heating of the gas via photoelectric heating effect and is therefore an important input parameter in chemical networks (e.g. \citealt{2007Jonkheid, 2009Woitke, 2010Kamp}). Since the degree of ionization of the PAH can be estimated by the UV field, the electron density and the disk temperature, we do not expect to see different behavior of the PAH ionization when including thermo-chemical networks in our disk model. 

We consider PAH with only two accessible ionization stages, the neutral and the singly charged cation. The neutral fraction, $f(0)$ is given by
\begin{equation}
\label{eq:f0}
f(0) = [1+\gamma_0]^{-1}
\end{equation}
with $\gamma_0$ the ratio of the ionization rate over the recombination rate, which, adopting the classical expression, is 
is 
\begin{equation}
\label{eq:gamma}
\gamma_0 = 3.5 \times 10^{-6} N_c^{1/2} \frac{G_0T^{1/2}}{n_e}.
\end{equation}

Here, the constant $3.5 \times 10^{-6}$ is derived from the probability that a colliding electron sticks \citep{2005Tielens}. The MCMax code calculates the neutral fraction $f(0)$ in each grid cell in the disk. For all grid cells with neutral fractions between $f(0)= 0$ (ionized) and $f(0) = 1$ (neutral), MCMax linearly interpolates between the neutral and ionized PAH emission profiles.

Figure \ref{fig:f0_gamma} shows the relation between the neutral fraction and $G_0 T^{1/2}/n_e$ for different PAH sizes. We find that most PAHs are emitting in the physical range $G_0 T^{1/2}/n_e \sim10^3 - 10^6$ in our disk models. Therefore, our two charge model covers the physical range of our interest.

The ultimate goal of our modeling is to understand the range of band strength ratios $I_{6.2}/I_{11.3}$ for the four transitional disks (HD\,97048, HD\,169142, HD135344\,B, and Oph IRS 48) and to learn what differences in disk properties are responsible for neutral versus ionized PAH emission. In the next section we first describe a benchmark model and explore how PAHs behave as a function of the relevant parameter space. After that, we present PAH models of four transitional disks, which fit the $I_{6.2}/I_{11.3}$ ratio by combining contributions from neutral PAHs in the dusty outer disk and ionized PAHs in an optically thin gas flow through the gap.

%-----------------------------------------------------------------------------------------
%					PAH MODELING
%-----------------------------------------------------------------------------------------

\section{Benchmark model}
\label{sec:results}
We study the behavior of the degree of ionization, the PAH luminosity, and the radial dependence of the PAH charge state throughout the disk. We focus on PAH emission from self-consistent disks models, which include high optical depth media. In this approach, the vertical density structure of the gas is set by hydrostatic equilibrium, and dust properties, such as scattering, grain sizes, and grain composition, are important input parameters for the radiative transfer. This allows us to evaluate how the charge state of PAH molecules in the disk translates to the spectrum. We start with a simple hydrostatic flaring disk model for which we adopt the stellar properties of HD\,97048. All relevant star and disk parameters, as well as the results of the PAH characteristics, are given in Table \ref{tab:fourmodels}. 

We assume a dust composition of 80\% silicate and 20\% amorphous carbon. The dust composition with reference to the optical constants is 12\% MgFeSiO$_4$ \citep{1995Dorschner}, 34\% Mg$_2$SiO$_4$ \citep{1996HenningStognienko}, 32\% MgSiO$_3$ \citep{1995Dorschner}, 2\% NaAlSi$_2$O$_6$ \citep{1998Mutschke} and 20\% C \citep{1993Preibisch}. The shape of our particles is irregular and approximated using a distribution of hollow spheres (DHS, \citealt{2005Min}). A vacuum fraction of 0.7 has been used, which breaks the perfect symmetry of the  spherical compact grain. Though we do not focus on detailed feature shapes  this choice therefore does not influence our results.

The dust size range is from $a_{min}= 1$ $ \upmu$m up to $a_{max}=1$ mm and follows a power-law distribution with $a_{pow}=-3.5$. The surface density of the gas and dust decreases proportional to r$^{-1}$. Typically in the ISM, the fraction of the mass locked up in PAHs relative to the total dust mass is $4 \times 10^{-2}$ \citep{2007DraineLi}. A comparison with the PAH abundance in protoplanetary disks may be misleading because gas and dust in disks are expected to be processed. Given the uncertainties in abundances, we find that a PAH-to-dust mass fraction of  $5.0\times10^{-4}$ fits the PAH luminosity of HD\,97048. We use a single size PAH molecule with $N_{C}=100$, which has a radius of $a_{PAH}=6$ {\AA}.  The assumption of PAH-size may be arbitrary because the ionization depends weakly this choice. Though this paper only intends to demonstrate the effect of ionization, and the choice of a typical PAH of $N_{C}=100$ therefore seems justified.

\begin{figure*}[htbp]
\centering
\includegraphics[width=0.9\columnwidth]{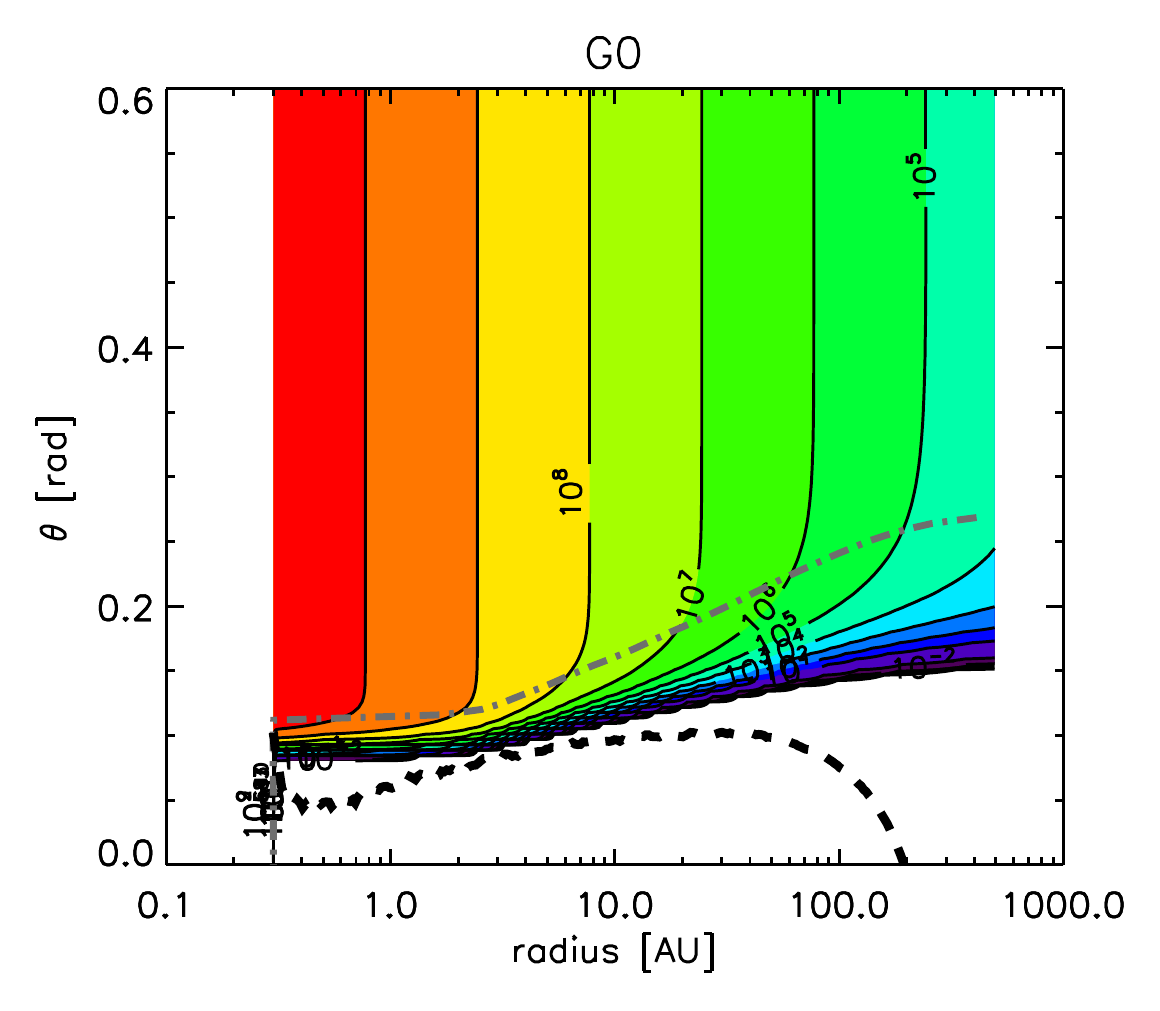}
\includegraphics[width=0.9\columnwidth]{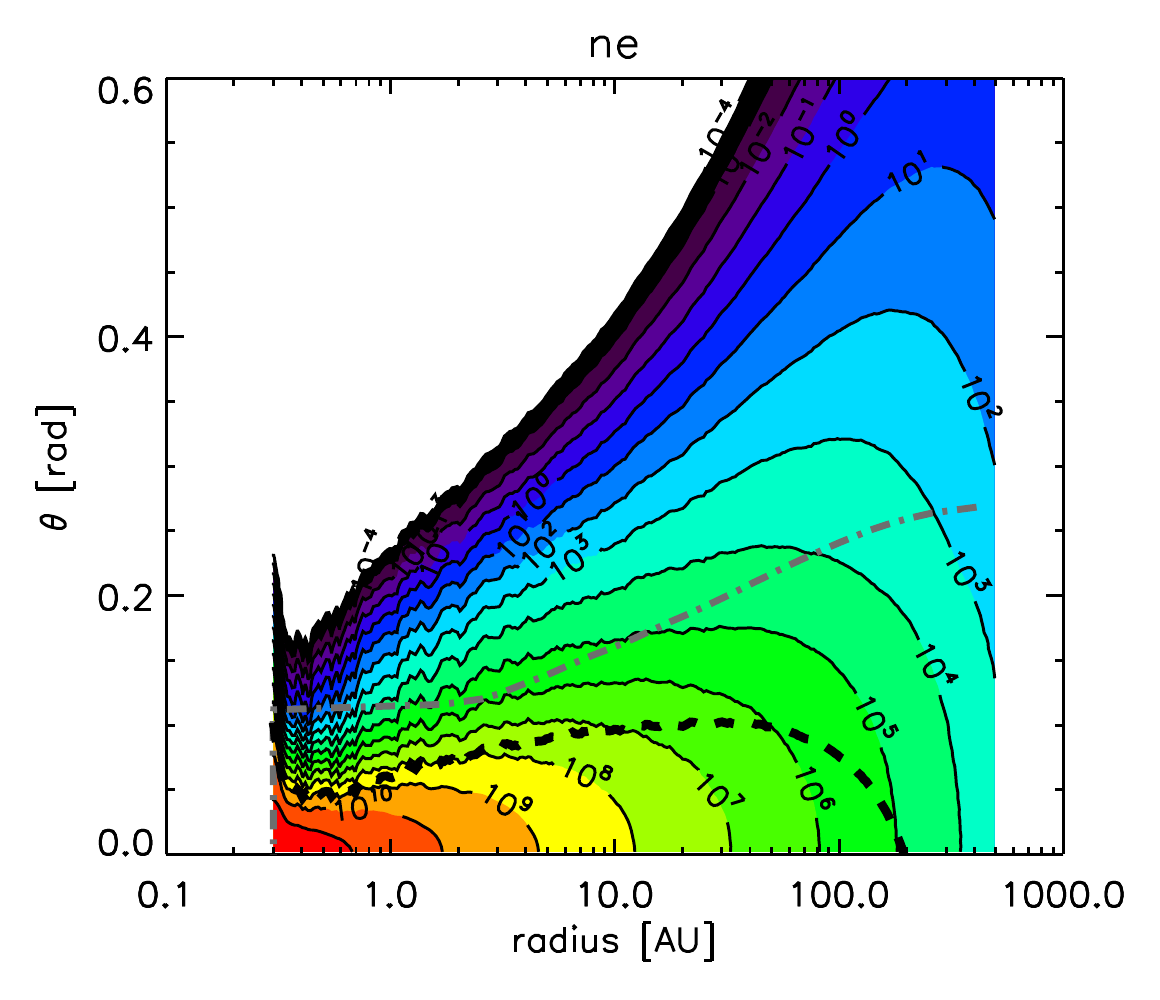}
\includegraphics[width=0.9\columnwidth]{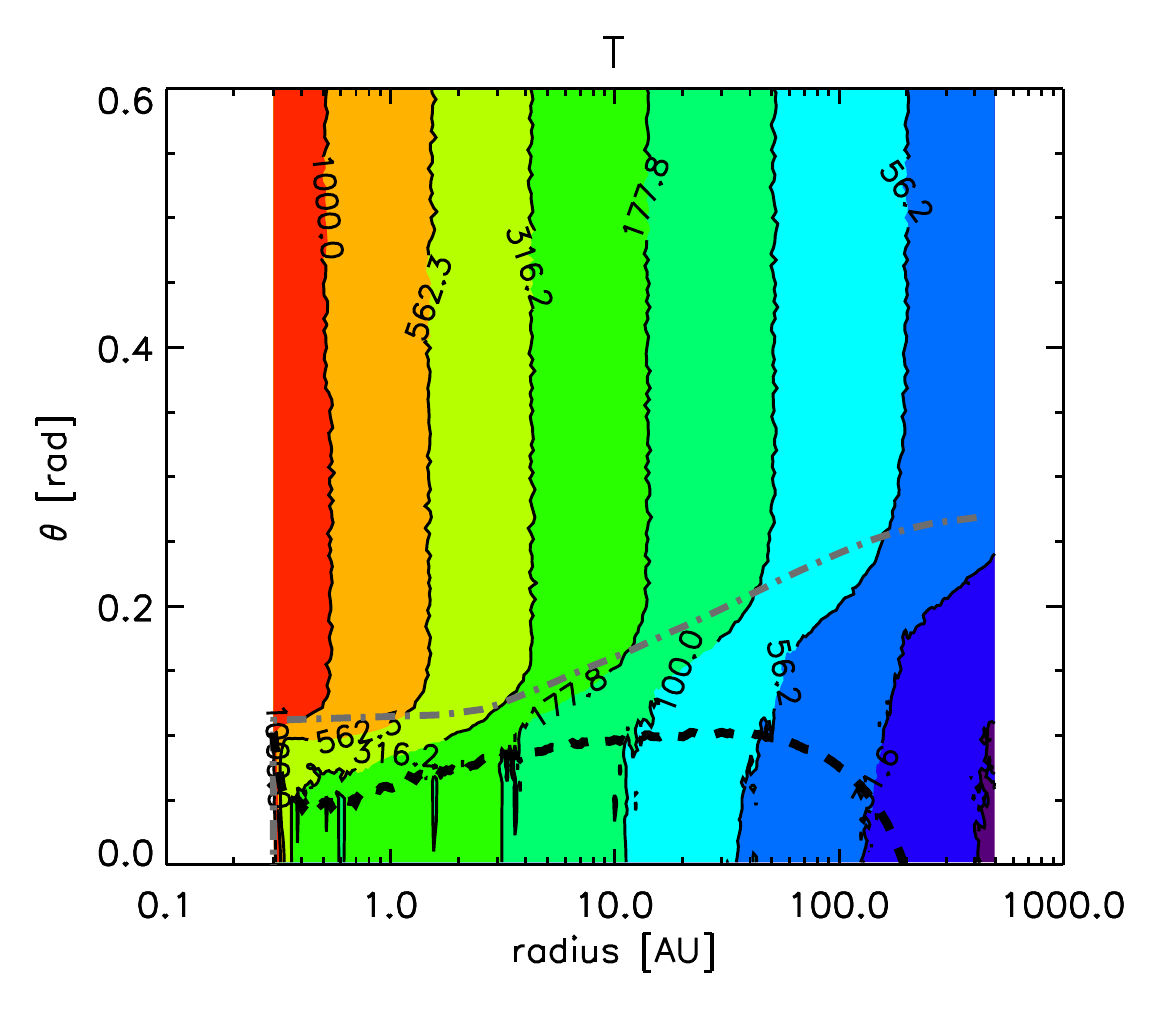}
\includegraphics[width=0.9\columnwidth]{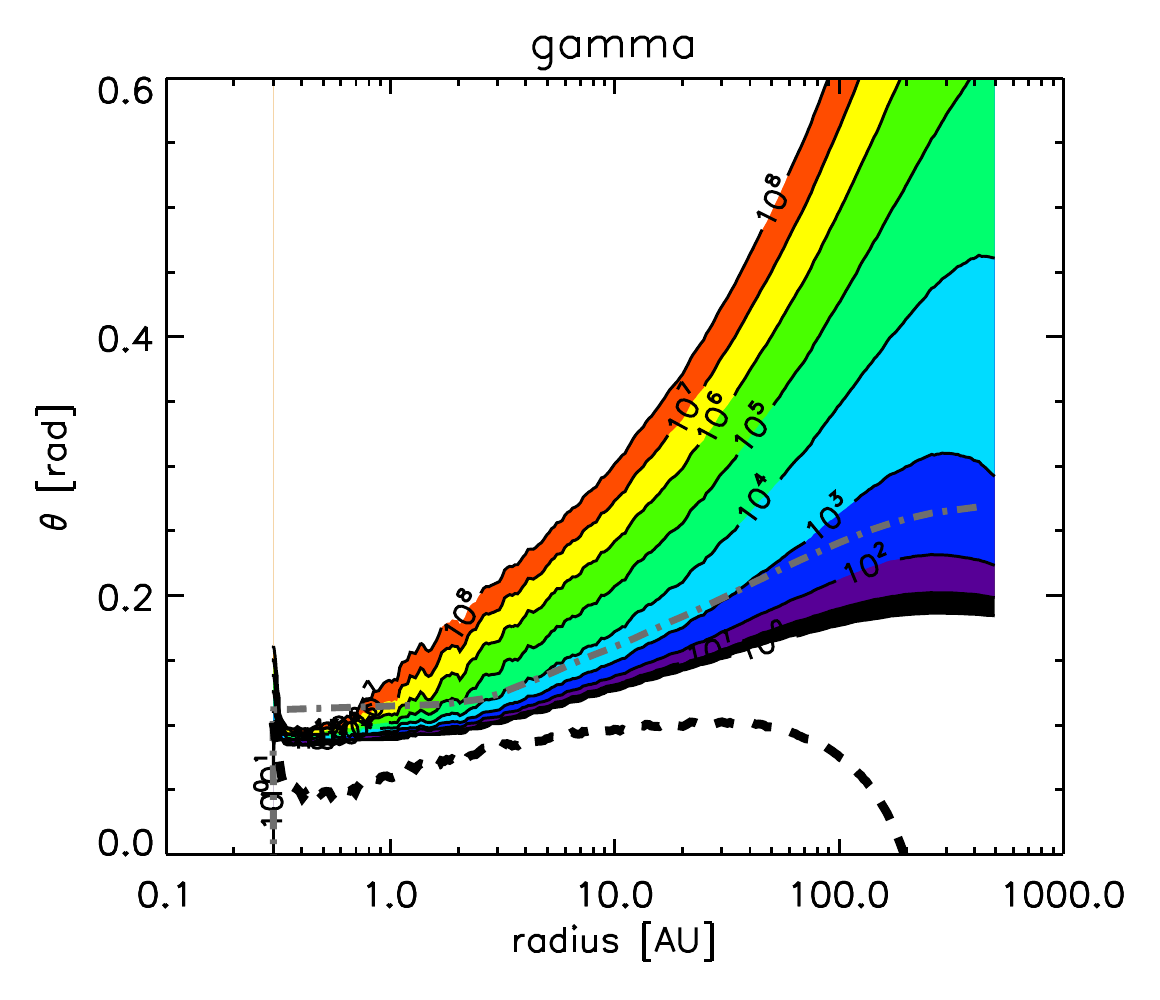}
\includegraphics[width=0.9\columnwidth]{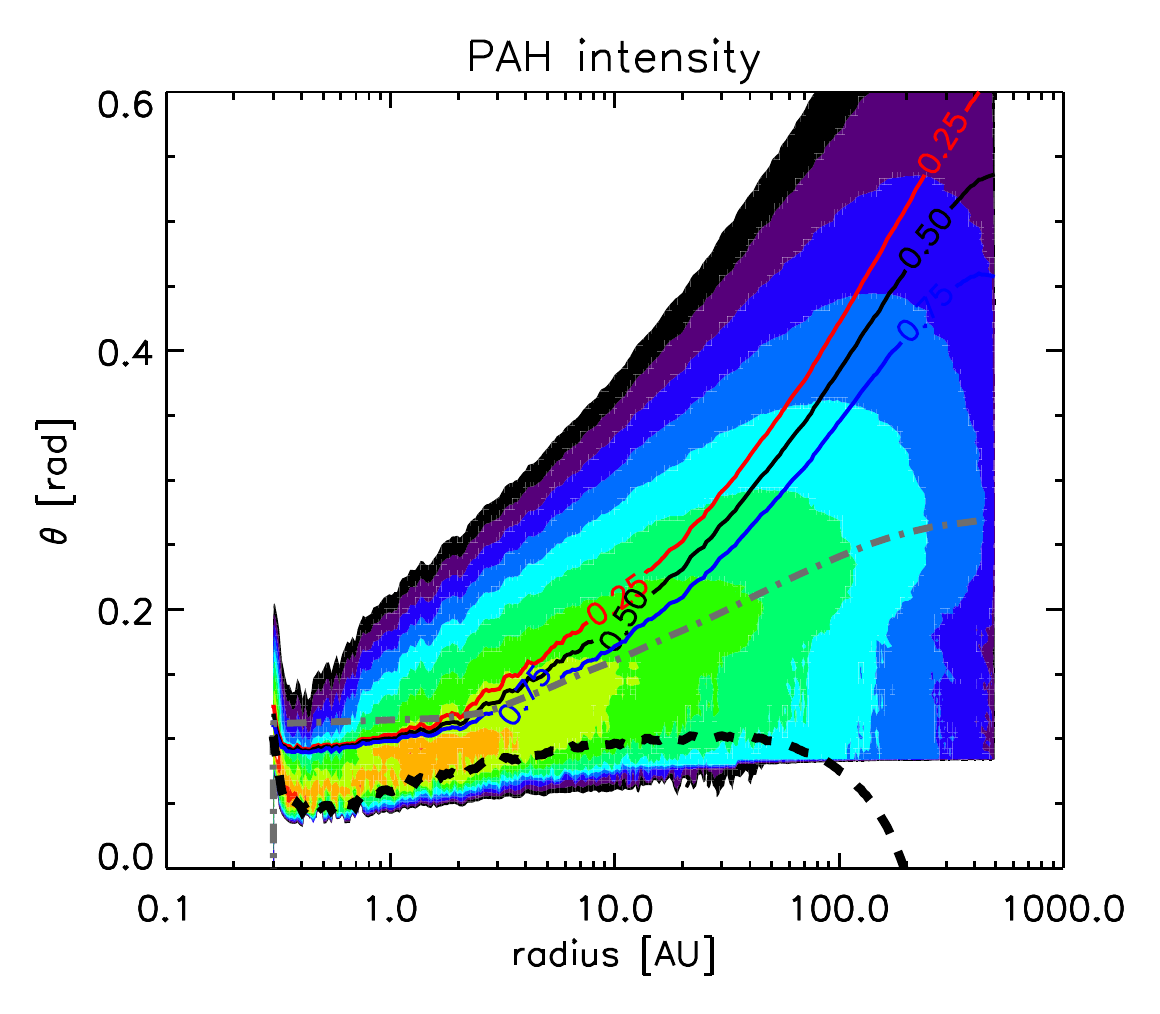}
\includegraphics[width=0.9\columnwidth]{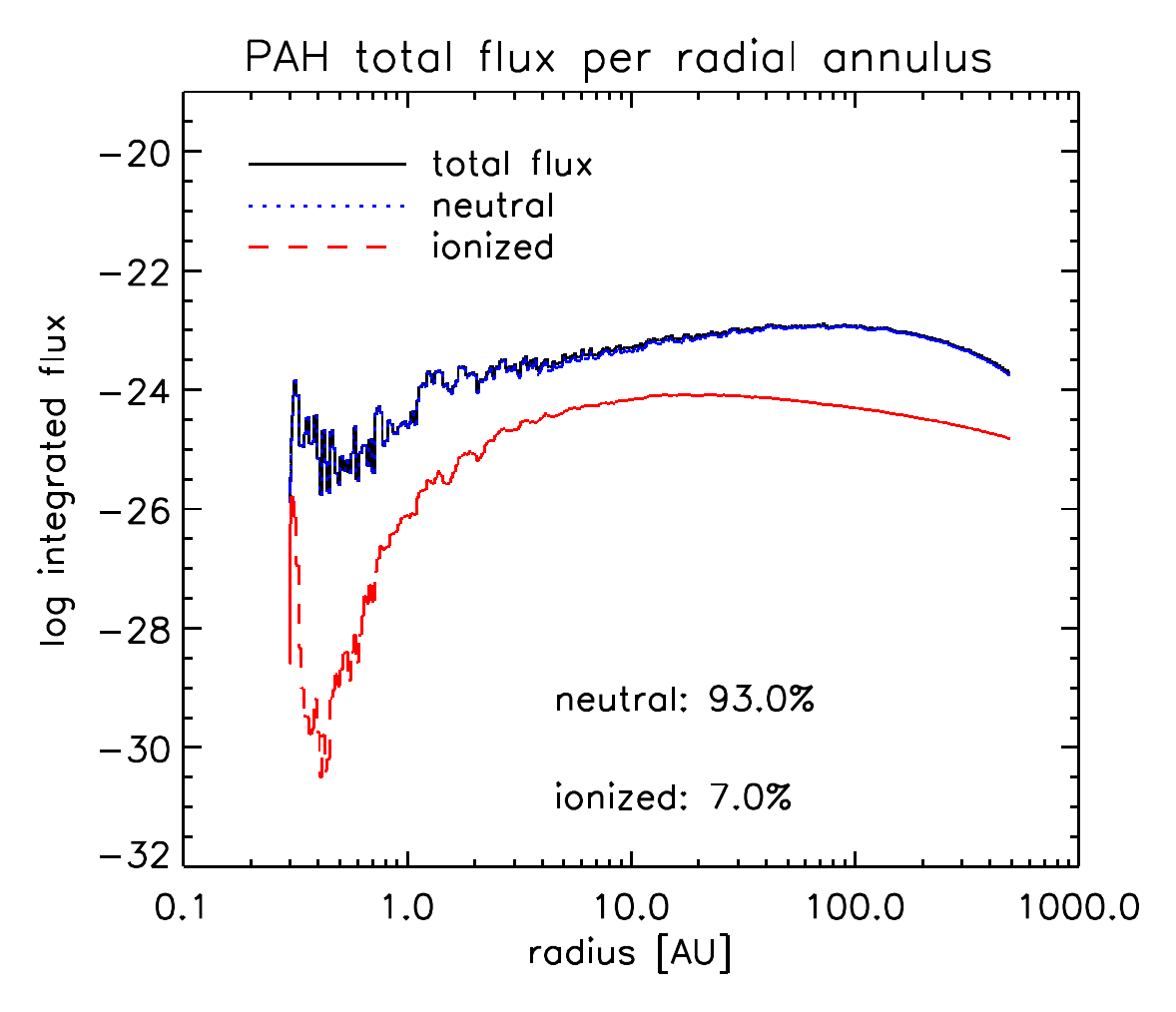}

\caption{ \label{fig:diagnosticplots}Diagnostic plots of benchmark model. The black dashed line is the vertical $\tau_{\mathrm{MIR}}=1$ surface at 10 $\upmu$m, the grey dashed-dotted line gives the radial $\tau_{UV}=1$ surface at 0.1 $\upmu$m. \textbf{Upper left:} the strength of the UV field $G_0$. \textbf{Upper right:} the electron number density $n_{e}$. \textbf{Middle left:} the temperature distribution in the disk. \textbf{Middle right:} the ionization parameter $\gamma$ given by $G_0 T^{1/2}/  n_e$. \textbf{Bottom left:} the PAH intensity of a 2D cut through the disk.  This diagnostic plot shows the origin of the PAH flux in the disk, i.e. the contributions of PAH emission to the final spectrum. The intensity of the PAH flux is shown in logarithmic scale (arbitrary scaling), where each colour spans an order of magnitude. The red, black and blue solid lines give the locations where the neutral fractions $f(0)$ are respectively 0.25, 0.5, 0.75. The neutral fractions are computed by equations \ref{eq:f0} and \ref{eq:gamma} in each grid cell of the disk. PAHs in the surface of the disk are largely ionized, while PAHs in the mid-plane are neutral. \textbf{Bottom right:} the intensity of PAH emission per radial annulus (arbitrary units). The blue and red lines give the neutral and ionized contributions. 93\% of the PAH molecules in the benchmark flaring disk model are neutral. The contribution from neutral PAHs dominate irrespective of the radial location in the disk. The noise is due to photon statistics.  The height $\theta$ relates to Z by, $\theta = \tan(Z/r)$. }
\end{figure*}

\begin{figure*}[t]
\includegraphics[width=\columnwidth]{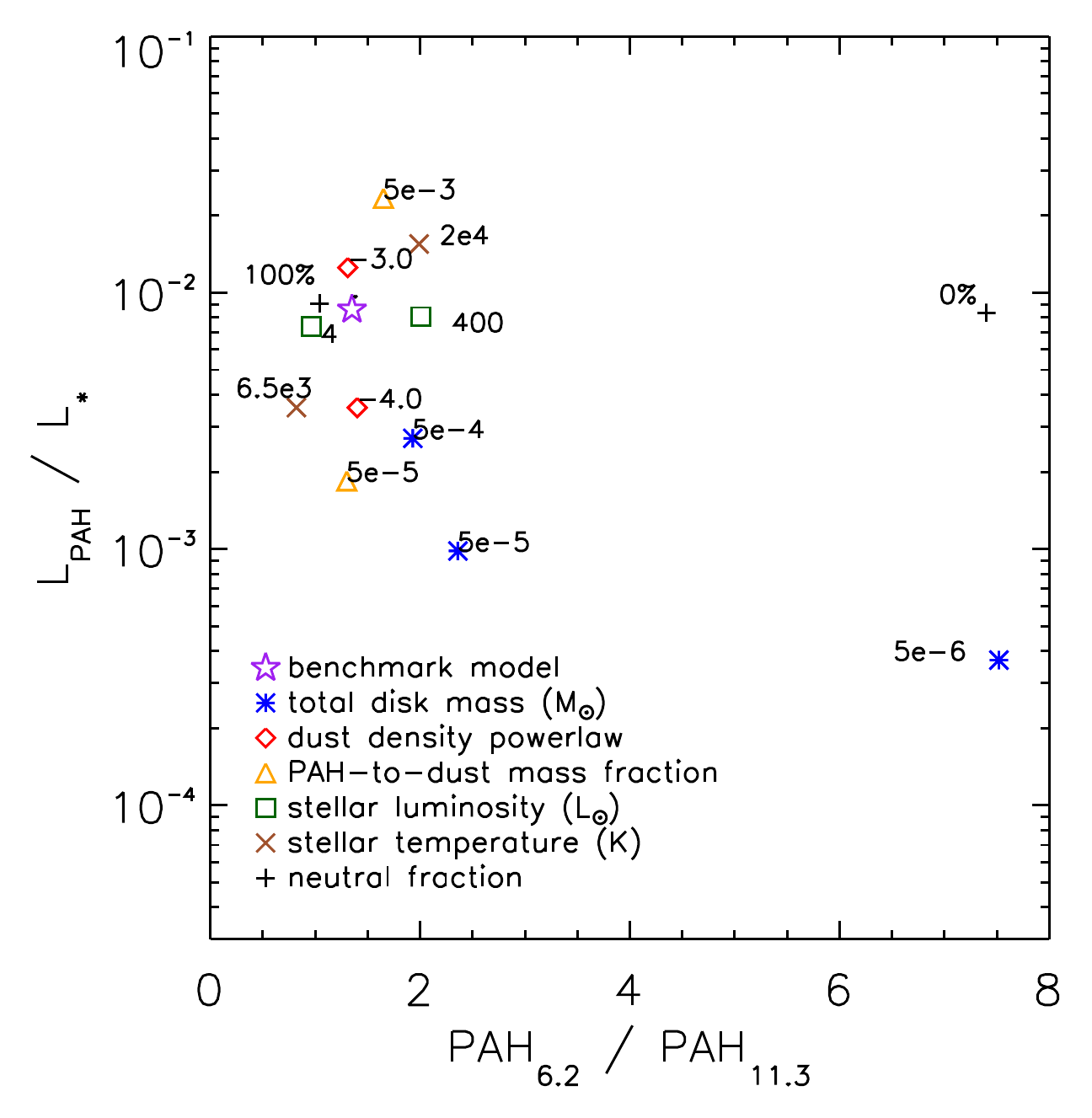}
\includegraphics[width=\columnwidth]{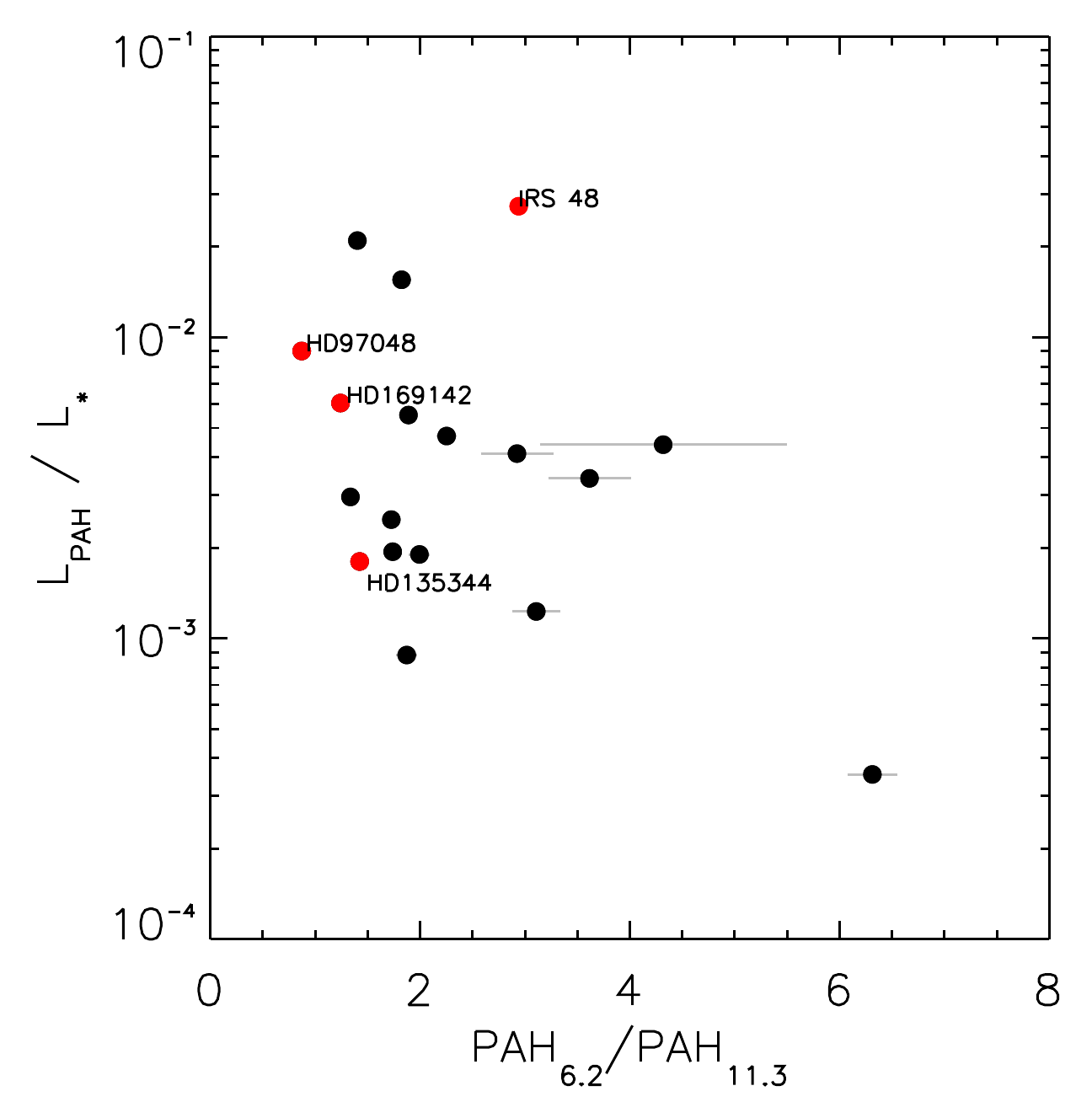}

\caption{\label{fig:parametergrid}PAH luminosities ($L_{PAH}/L_{*}$) compared to the $I_{6.2}/I_{11.3}$ ratio for the benchmark modeling grid (left) and the observations of Herbig stars (right). Several parameters of the benchmark model are varied and their effect to the PAH luminosity and $I_{6.2}/I_{11.3}$ ratio is shown. The main results of the benchmark modeling grid study is that in general the PAH neutral fraction is high and ionized PAHs originate from optically thin, low density environments. The red dots show the objects studied in this paper. The parameters of all the objects shown in this right plot are listed in Table \ref{table_sed_parameters}.}
\end{figure*}

Figure \ref{fig:diagnosticplots} presents six diagnostic plots of the benchmark model. The black dashed-dotted line gives the radial $\tau_{UV}=1$ surface at 0.1 $\upmu$m. This is the location where UV-photons from the central star are scattered inward in the disk. The gray dashed line gives the vertical $\tau_{MIR}=1$ surface at 10 $\upmu$m. Above this line, MIR photons can escape the disk. Therefore, it shows the region of the disk where the PAH emission originates. The upper four plots give, respectively, the strength of the UV field $G_0$, the distribution of the electron number density $n_{e}$, the temperature $T$ in the disk, and the ionization parameter $\gamma$ given by $G_0 T^{1/2}/ n_e$. The ionization parameter $\gamma$ is a representation of the physical conditions, which control the ionization balance of PAHs. Thus, at the radial $\tau_{UV}=1$ surface between 10 AU and 100 AU, $\gamma \sim10^3-10^4$. The bottom left plot shows the PAH intensities (arbitrary scaling) responsible for the observed emission.  The PAH intensity shows the relative contributions of PAH emission to the final spectrum. The region where the PAH contribution is highest lies between the radial $\tau_{UV}=1$ and vertical $\tau_{MIR}=1$ surfaces. The red, black and blue solid lines give the locations, where the neutral fractions $f(0)$ are, respectively 0.25, 0.5 and 0.75. The neutral fractions are computed in each grid cell of the disk using equations \ref{eq:f0} and \ref{eq:gamma}. From this, we can see that most of the PAH emission originates in values below the $f(0)=0.75$ line and, thus, have a high neutral fraction. The bottom right plot shows the intensity of the integrated PAH emission per radial annulus (arbitrary units). The PAHs between $\sim$20--200 AU contribute most to the final spectrum. The blue and red lines give the decomposed neutral and ionized contributions. Of the PAH molecules, 93\% in the benchmark flaring disk model are neutral. 

An important insight from the benchmark model is that neutral PAHs dominate the emission at all radii in the disk. There is, thus, no gradient in the neutral fraction of PAHs as a function of radius in an optically thick disk. Only the surface layer of the disk produces ionized PAH emission but that contribution is $\sim$1--2 orders of magnitude smaller than the contribution from neutral PAHs from deeper layers in the disk. The effect of a higher UV field does not translate to a higher degree of ionization because the electron density is also higher closer to the star. 

The PAH luminosity of the benchmark model is $L_{PAH}/L_{*}=8.40\times10^{-3}$ and band strength ratio $I_{6.2}/I_{11.3}=1.35$. The results are shown on the top of Figure \ref{fig:parametergrid}, where a direct comparison can be made to the observations of our sample of Herbig stars, as shown on the bottom plot. As a next step, we investigate the effects of varying the star and disk parameters of the benchmark model on $L_{PAH}$ and $I_{6.2}/I_{11.3}$. We discuss the parameters which have a significant effect on the charge state of the PAHs.

\begin{table*}[htdp]
\caption{\label{tab:fourmodels}Characteristics of benchmark model and models of four transitional disks. }
\begin{center}
\begin{tabular}{l l l c c c c c}
\hline
\hline
\multicolumn{2}{c}{Parameter }							& Unit	&Benchmark 			&	HD\,97048		&	HD\,169142		&	HD\,135344\,B 			&	Oph IRS 48    \\
\hline

Stellar temperature		&		$T_{*}$				& K		 		&10 000 			&  10 000 	 			  &   8200     			 &  6590				       & 9000  \\	
Stellar luminosity		&		$L_{*} $				& L$_{\odot}$		& 40 				& 40 		   			 &    15.3    			 &  8.26				       &14.3	  \\
Stellar radius			&		$R_{*}$				&R$_{\odot}$		&	2.12 			&   2.12 				  &  	1.94			 	  &  2.20				       &2.30	  \\
Stellar mass			&		$M_{*}$				&M$_{\odot}$		 &	2.50 			&  2.50 	   			&    2.28    				 &  3.30				       &2.25  \\
inclination				&		$i$					&$^{\circ}$		&45				&43					&13					&11						&48\\
Distance				&		$d$					&pc 				&158 			&  158				   &  145			      	 &  140				       &120  \\
\vspace{0.001mm}\\
Carbon atoms in PAH	&		$N_{C}$				&	\dots		&  100 				&100     				&    100     				&    100     					& 100 \\
Silicate mass			&		$M_{Si}$				&  M$_{\mathrm{dust}}$	&0.8 					&0.8 		  	 		&  0.8 	       			&   0.8 	     				& 0.8 	 \\
Carbon mass			&		$M_{C}$				& M$_{\mathrm{dust}}$	& 0.2 				&   0.2 		 	 	& 0.2 	      			&  0.2 	      				&  0.2 	   \\
Min dust size			&		 $a_{min}$			& $\upmu$m	&  1 					&   0.5  				&  0.5       				&   1      					&  0.1\\
Max dust size			&		 $a_{max}$			&  mm 		&   1					&     1				& 1				        & 1					        & 1	 \\
Dust-size power-law index	&		$a_{pow}$			& 	\dots		& -3.5 				&    -3.5 				&  -3.5			       & -4.0      					  & -4.0 \\

\vspace{0.001mm}\\
%UV field				&		$<G_0>$				& u$_{hab}$	&5.27$\times$10$^7$	&	1.69$\times$10$^7$	&	9.51$\times$10$^6$		&	 1.09$\times$10$^6$	& 7.03$\times$10$^7$	\\
%Electron density 		& 		$<{n_e}>$			& 	cm$^{-3}$	&1.02$\times$10$^7$	&	2.98$\times$10$^7$	 &	7.57$\times$10$^5$		&	1.68$\times$10$^5$		 &	7.37$\times$10$^3$ \\
%Temperature 			& 	$<\mathrm{T}_{PAH}>$ 		&	K		 &121				&	94 				&	131					&	162						&	164 \\
%Final, ionization parameter	&	$<\gamma>$				& 	\dots		&2.38$\times$10$^5$	&	1.00$\times$10$^5$	&		8.37$\times$10$^5$	 &		2.73$\times$10$^4$	&		8.20$\times$10$^5$	 \\
Total, PAH neutral fraction		&	$<f(0)_{total}>$	 	&	\dots		&0.93				&	0.97				 &	0.92					 &	0.91					&	0.54		\\
Total, PAH ratio	&	$I_{6.2}/I_{11.3}$				& 	\dots		& 1.35				&	1.27		 		&		1.36				&		1.42					&		3.30		\\
Total, PAH luminosity	&	$L_{PAH,total} / L_{*}$		& 	\dots		&$8.40\times10^{-3}$	&	$9.13\times10^{-3}$&		$6.24\times10^{-3}$&		$3.45\times10^{-3}$	&		$1.49\times10^{-2}$\\

\vspace{0.001mm}\\
Outer disk, PAH neutral fraction&	$<f(0)_{disk}>$	 		&	\dots			&0.93				&	0.97				 &1.00				 &	1.00					&	0.98		\\
Outer disk, PAH luminosity&	$L_{PAH,disk} / L_{PAH,total}$& 	\dots			&1					&	0.95				&	0.67				&	0.27					&	0.09		\\
Outer disk, gas mass	&		$M_{gas,disk}$			& M$_{\odot}$		& $5.0\times10^{-2}$ 	&  $5.0\times10^{-2}$	 &  $ 8.0\times10^{-3}$     	& $1.0\times10^{-2}$      		& $3.0\times10^{-3}$  \\		
Outer disk, dust mass	&		$M_{dust,disk}$		&  M$_{\odot}$		&$5.0\times10^{-4}$ 	&   $5.0\times10^{-4}$	&  $8.0\times10^{-5}$    	& $1.0\times10^{-4}$    		  & $3.0\times10^{-5}$  \\ 
Outer disk, PAH mass 	&		$M_{PAH,disk}$		& M$_{\odot}$		&$2.5\times10^{-7}$ 	&  $2.5\times10^{-7}$ 	 & $  4.0\times10^{-8}$ 	&    $5.0\times10^{-8}$  		 &  $1.5\times10^{-8}$  \\
Outer disk, PAH-to-dust mass fraction &	$f_{PAH,disk}$		& \dots			&$5.0\times10^{-4}$ 	&  $5.0\times10^{-4}$ 	 & $  5.0\times10^{-4}$  	&  $  5.0\times10^{-4}$		 &  $5.0\times10^{-4}$   \\
Outer disk, inner radius  	&		$R_{in,disk}$			&AU				&0.3 					& 34		   			 &   23      				& 30      		 			 & 63 \\
Outer disk, outer radius 	&		$R_{out,disk}$			&AU				&500 				& 500   				 & 235        			&  235       					& 235 \\

%Outer disk, ionization parameter	&	$<\gamma>$		& 	\dots			&2.38$\times$10$^5$	&	1.00$\times$10$^5$	&	3.33$\times$10$^3$	 &		2.81$\times$10$^1$	&		1.13$\times$10$^5$	 \\
%UV field				&		$<G_0>$				& u$_{hab}$		&5.27$\times$10$^7$	&	1.69$\times$10$^7$	&3.10$\times$10$^5$	&	 3.81$\times$10$^3$	& 8.12$\times$10$^4$	\\
%Electron density 		& 		$<{n_e}>$			& 	cm$^{-3}$		&1.02$\times$10$^7$	&	2.98$\times$10$^7$	 &9.04$\times$10$^5$	&	5.50$\times$10$^5$		 &	2.78$\times$10$^4$ \\
%Temperature 			& 	$<\mathrm{T}_{PAH}>$ 		&	K			 &121				&	94 				&111				&	88					&	102 \\

\vspace{0.001mm}\\
Disk gap, PAH neutral fraction	&	$<f(0_{gap})>$	 		&	\dots		&\dots				&	0.21				 &	0.07				 &	0.87					&	0.52		\\
Disk gap, PAH luminosity	&	$L_{PAH,gap} / L_{PAH,total}$& 	\dots		&\dots				&	0.05				&	0.33				&	0.73					&	0.91			\\
Disk gap, gas mass		&		$M_{gas,gap}$			& M$_{\odot}$		& \dots 			 &  $1.0\times10^{-7}$	 &  $1.4\times10^{-8}$     	&$ 9.5\times10^{-8}$     		&$1.9\times10^{-6}$ \\		
Disk gap, dust mass		&		$M_{dust,gap}$		&  M$_{\odot}$		&\dots 			 &  $1.0\times10^{-9}$	&  $1.4\times10^{-10}$    	& $9.5\times10^{-10}$     		 &$ 1.9\times10^{-8}$  \\ 
Disk gap, PAH mass 	&		$M_{PAH,gap}$		&M$_{\odot}$		&  \dots			& $3.8\times10^{-11}$ 	 & $  5.8\times10^{-12}$ 	 &    $3.8\times10^{-11}$    	&  $7.7\times10^{-10}$  \\
Disk gap, PAH-to-dust mass fraction&		$f_{PAH,gap}$	& \dots			& \dots			&   $4.0\times10^{-2}$  	&  $ 4.0\times10^{-2}$  	 &   $ 4.0\times10^{-2}$ 		&  $4.0\times10^{-2}$    \\
Disk gap, inner radius  	&		$R_{in,gap}$			&AU				&\dots 			&2.5		   			 &   0.3      				& 0.3      		 			 & 0.3 \\
Disk gap, outer radius 	&		$R_{out,gap}$			&AU				&\dots  			& 34   				 & 23        				&  30       					& 63 \\

%Disk gap, ionization parameter	&	$<\gamma>$		& 	\dots		&	\dots				&	\dots				&		5.11$\times$10$^6$	 &	3.91$\times$10$^4$	&	8.62$\times$10$^5$	 \\
%UV field				&		$<G_0>$				& u$_{hab}$	&\dots				&	\dots				&	5.67$\times$10$^7$		&	 1.56$\times$10$^6$&	 7.44$\times$10$^7$	\\
%Electron density 		& 		$<{n_e}>$			& 	cm$^{-3}$	&\dots				&	\dots				 &	5.19$\times$10$^2$		&	2.22$\times$10$^3$	&	6.18$\times$10$^3$ \\
%Temperature 			& 	$<\mathrm{T}_{PAH}>$ 		&	K		 &\dots				&	\dots				&	229					&	194				&	168 \\

 \hline
\end{tabular}
\end{center}
\end{table*}%

\subsection{Stellar properties}

We varied the stellar temperatures and luminosities of the benchmark model. A decrease in stellar temperature and luminosity lowers the neutral fraction because the UV field $G_{0}$ decreases and fewer UV photons are available to excite the PAHs. Therefore, the $I_{6.2}/I_{11.3}$ ratio goes down. For a higher luminosity and/or stellar temperature, the UV field $G_{0}$ and, consequently, the $I_{6.2}/I_{11.3}$ ratio are higher. Figure \ref{fig:parametergrid} shows that our chosen range of stellar temperatures and luminosities result in band strength ratios $I_{6.2}/I_{11.3}$ between $\sim1-2.2$. This range covers 11 of the 18 Herbig stars shown on the right of Figure \ref{fig:parametergrid}. None of the other 7 objects with higher  $I_{6.2}/I_{11.3}$ ratios have temperatures and/or luminosities higher than 20 000 K and/or 400 $L_{\odot}$. Thus, for these objects, the stellar temperature and luminosity cannot explain the high band strength ratio  $I_{6.2}/I_{11.3}$. 

For stars with a lower temperature and/or luminosity, the UV density decreases and the contribution from PAHs heated by multi-photon events is lower. As PAHs heated by multi-photon events have a higher temperature, they also have a slightly higher  $I_{6.2}/I_{11.3}$ ratio and vice versa.  In the models with a lower stellar luminosity and temperature all the PAHs in the disk are neutral, although the $I_{6.2}/I_{11.3}$ ratio is slightly lower ($\sim0.1-0.2$) than the 100\% neutral benchmark model (see Figure \ref{fig:parametergrid}). Thus, this difference is a result of the fraction of PAH emission heated by multi-photon versus single-photon events. However the difference is small and unimportant compared to the effect of ionization.

\subsection{Grain properties}

To test the effect of the grain size distribution, we change the dust density power-law to $a_{pow}=-3.0$ and $a_{pow}=-4.0$. For a relatively lower abundance of small grains ($a_{pow}=-3.0$), the average dust opacity is lower and the disk is less opaque.  As a result, the PAH luminosity increases, since the optical depth in the disk is lower, and the region where PAHs can get excited is larger.  For a higher abundance of small grains ($a_{pow}=-4.0$), the PAH luminosity decreases, since the optical depth in the disk is higher and the region where PAHs can get excited is smaller. The size distribution of dust grains has a strong effect on the PAH luminosity through the dust optical depth in the UV. If small dust grains are depleted, then the UV opacities in the disk are dominated by PAHs and a small amount of PAHs can give rise to a strong PAH luminosity. The effect on the degree of ionization is negligible since the ionization parameter $\gamma$ and the neutral fraction $f(0)$ change accordingly to the optical depth and the fraction of ionized to neutral PAHs stays roughly similar. 

\subsection{PAH properties}

Since evolutionary processes may affect the PAH composition in the disk, it is difficult to constrain PAH sizes and abundances. In our benchmark model, we assume a PAH mass of $5.0\times10^{-4}$ M$_{\mathrm{dust}}$. For an order of magnitude higher and lower PAH mass fraction, the PAH luminosities change by factors of 2.7 and 0.21 respectively. The PAH mass fraction may then explain the wide spread in PAH luminosity among Herbig stars ($\sim$3 orders of magnitude, see Table \ref{table_sed_parameters}) 

Figure \ref{fig:parametergrid} shows a small shift in the $I_{6.2}/I_{11.3}$ ratio for factor of 10 higher or lower PAH-to-dust mass fraction. This is also because the PAH-to-electron abundance is then altered, which changes the amount of electrons per PAH molecule available for recombination. Because PAH sizes in disks are expected to be in the order of  $N_{C}\sim50-150$ \citep{2008Tielens}, the difference in PAH sizes have a very small effect on the final spectra because the neutral fraction depends only on the square root of the PAH size in our ionization model (equation \ref{eq:gamma}). 
%\todo{Check with Michiel: I have models with Nc=10 and Nc=1000. However, I do not completely understands what is happening since the I\_6.2/I\_11.3 ratio of both models go down (1.30 and 1.42). Thus another effect (perhaps PAH heating and cooling) seems to balance the effect of ionization. }

\subsection{Disk mass}

We test the effect of lower disk masses on the PAH characteristics by decreasing the total disk mass from $5\times10^{-2}$ M$_{\odot}$, which is four orders of magnitude down to $5\times10^{-6}$ M$_{\odot}$. For a less massive disk, the disk vertical height and, consequently, the fraction of the stellar luminosity captured by the disk are lower. Therefore, the PAH luminosity drops with decreasing disk mass, even when the disk is still optically thick.

The charge state increases when the mass of the disk becomes so small (M$_{disk}\lesssim 5\times10^{-5} $ M$_{\odot}$ ) that the disk becomes optically thin. At that point, the UV field $G_0$ is high even in the mid-plane of the disk, while the electron density $n_e$ is lower. As a result, the fraction of ionized PAHs in the benchmark model with a disk mass of $5\times10^{-6}$ M$_{\odot}$ is almost 100\% and the  $I_{6.2}/I_{11.3}$ ratio is 7.82. This low disk mass model and the Herbig star HD141569 have a similar $L_{PAH}$ and $I_{6.2}/I_{11.3}$ (see also Table \ref{table_sed_parameters}). This is consistent with our modeling result because HD\,141569 is in the transition of a gas-rich protoplanetary disk to a gas-poor debris disk and largely optically thin \citep{2003LiLunine, 2013Thi}. 

\subsection{Summary}
The primary goal of our study is to identify the most important trends in the behavior of $L_{PAH}$ and the $I_{6.2}/I_{11.3}$ ratio. We have presented a small parameter study around the benchmark model that investigates the behavior of PAH emission as a function of disk properties. We have covered the disk parameters which have a significant effect on the charge state and/or luminosity of the PAHs; the results are shown in Figure \ref{fig:parametergrid}. The main insights are the following. \textbf{1)} The neutral fraction is high ($f(0)\sim0.8-0.9$) for optically thick disks with properties typical for Herbig stars. \textbf{2)} The PAHs in an optically thick flaring disk are predominantly neutral at all radii. Thus the PAH neutral fraction shows no gradient with radius. \textbf{3)} The total PAH luminosity increases if the UV field is stronger if the PAH-to-gas mass fraction is higher and if the dust density is lower. \textbf{4)} The PAH ionization increases if the UV field is stronger and if the PAHs originate from low-density optically thin environments, where the electron density is low and the UV field is high.

\section{Transitional disks}

\label{sec:PAHfits}
 
In this section, we show that fitting the PAH spectra of four transitional disks requires a significant contribution of ionized PAHs in Ôgas flowsÕ through the gap. We perform quantitative fits to the PAH luminosities $L_{PAH}/L_{*}$ and band strength ratios $I_{6.2}/I_{11.3}$ of four transitional disks. We fit $I_{6.2}/I_{11.3}$ and $L_{PAH}/L_{*}$  as close as possible to the observed values with a maximum discrepancy of a factor of 2. The presented PAH models demonstrate that a physical disk solution predicting an enhanced contribution of ionized PAHs from low-density, optically thin regions is successful in explaining the observed spectra. It is also consistent with the available spatial information shown in Figure \ref{four_objects_PAHs}.

We adopt the disk structures of HD\,97048, HD\,169142, HD\,135344\,B, and Oph IRS 48 derived by \citet{2013Maaskant} (see Table \ref{tab:fourmodels}). All disks have large dust-depleted gaps and flaring outer disks. These large gaps are inferred from analysis of Q-band direct imaging. The absence of the 10- and 20-$\upmu$m silicate feature indicates that the gaps are radially very large and empty in small dust grains. All disks show a NIR excess indicative of a hot ($\sim$1500 K) component. To fit the SED at NIR wavelengths, an inner dust component has been added to the disk. For HD\,97048, this inner disk is optically thick, is in hydrostatic equilibrium and is located from 0.3 -- 2.5 AU. For HD\,169142, HD\,135344\,B, and Oph IRS 48, optically thin spherical dust halos are added between 0.1--0.3 AU to account for an even stronger excess of NIR emission. For the outer disks, we assume that the vertical scale height is set by hydrostatic equilibrium. For the dusty disk, we assume a radial dependence of the surface density $\Sigma \propto r^{-1}$. Therefore, the density is highest in the mid-plane close to the star. If the inner disk is optically thick (as for HD\,97048), the inner disk acts as a UV-shield and reduces the radiation field $G_0$ further out in in the disk.

 \begin{figure}[t]
\centering
\includegraphics[width=\columnwidth]{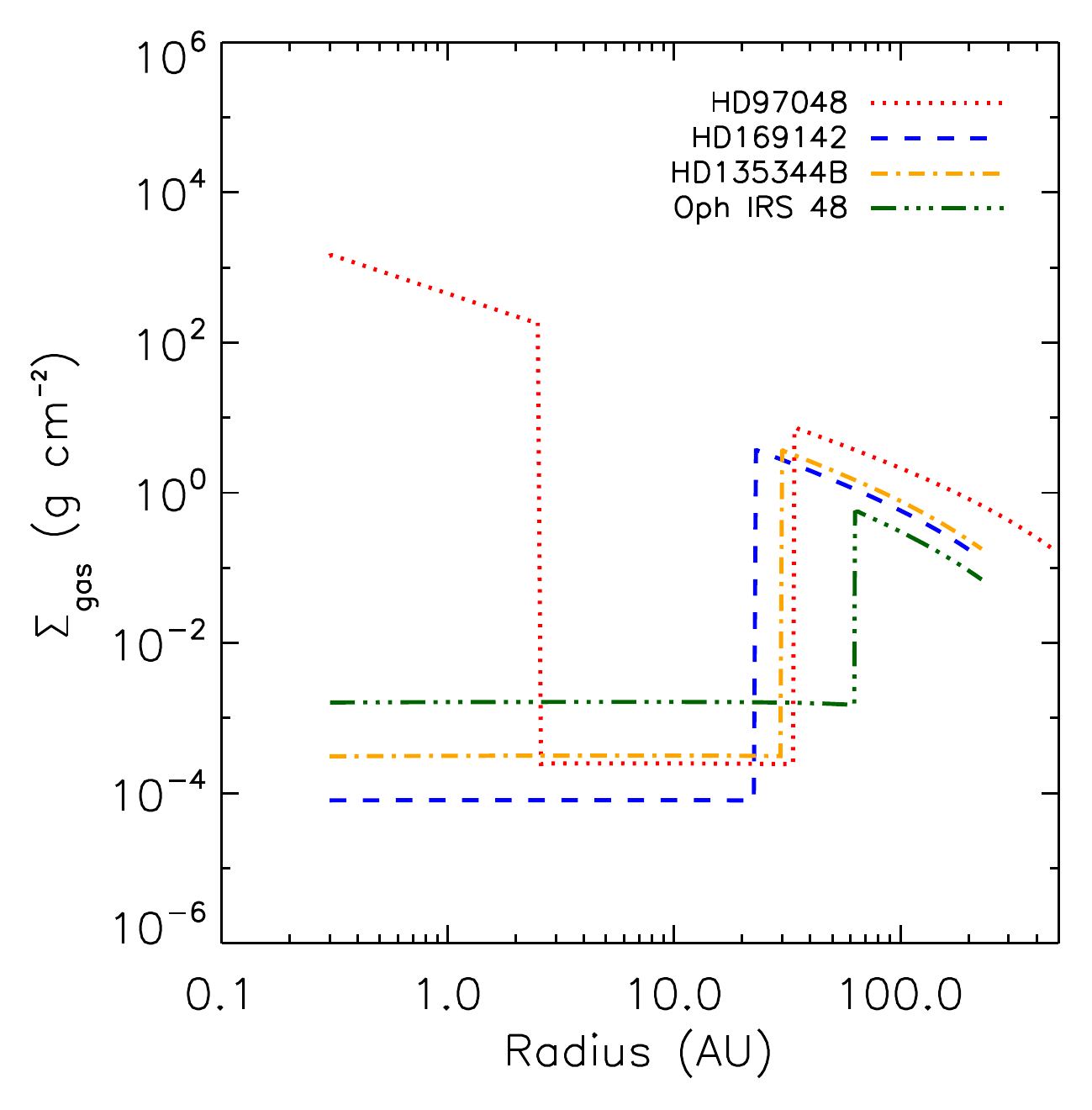}
\caption{ \label{fig:surf_dens}The gas surface densities $\Sigma_{gas}$ for the PAH models as a function of radius. Gas surface densities of $\Sigma_{gas}  \sim10^{-3}-10^{-4}$ g cm$^{-2}$ in the gaps are consistent with hydrodynamical planet formation models and represent gas flows through protoplanetary gaps. PAHs in these low-density gaps have a high ionization fraction.   }
\end{figure}

We use the spatial information from the VLT/VISIR FWHM profile, as shown in Figure \ref{four_objects_PAHs}, to understand the distribution of PAHs throughout the disk. However, this information is only taken into account qualitatively, as we refrain from doing a detailed fit of the FWHM profile since there are issues with the point spread function (PSF) of the VISIR measurements. In addition, the origin of the N-band continuum is not well understood since it can be produced by very small grains or dust from the inner and/or outer disk (see \citealt{2013Maaskant} for an extended discussion of this topic). A qualitative interpretation of the VLT/VISIR data is that the PAHs clearly come from larger scales than the N-band continuum for HD\,97048; thus, PAH emission originates in the outer disk (in agreement with \citealt{2006Lagage, 2006Doucet}). For HD\,169142, HD\,135344\,B and Oph IRS 48, the FWHM in the PAH feature is equal and even smaller than the continuum. If PAH emission would originate only from the outer disk, than the 11.3-$\upmu$m PAH feature would be resolved with respect to the FWHM of the continuum. Thus we interpret this as a significant contribution of PAH emission that comes from within the dust-depleted gap of the disk. 
\subsection{Adding low-density gas flows to the disk structure}

 \begin{figure*}[htbp]
\centering
\includegraphics[width=\columnwidth]{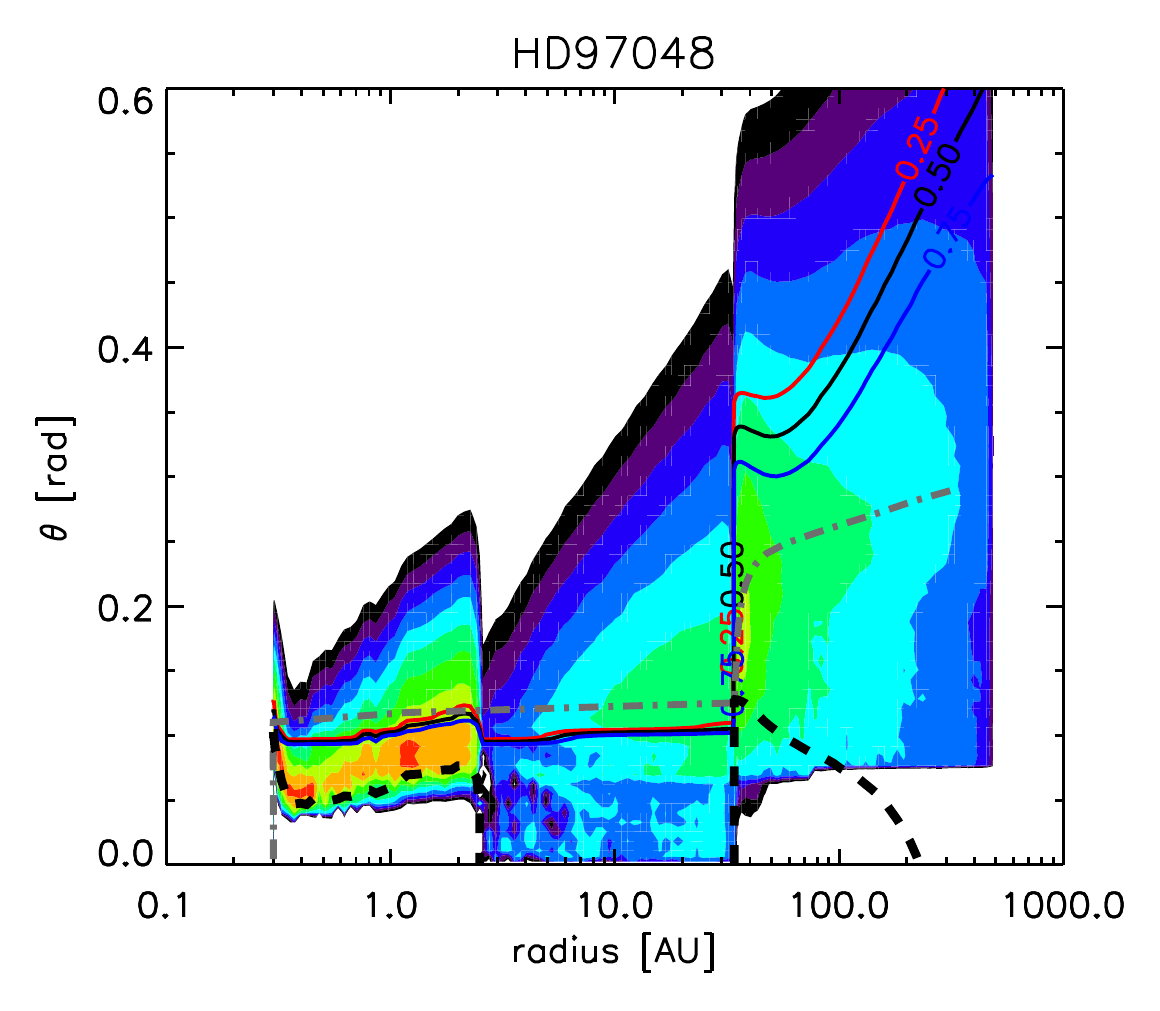}
\includegraphics[width=\columnwidth]{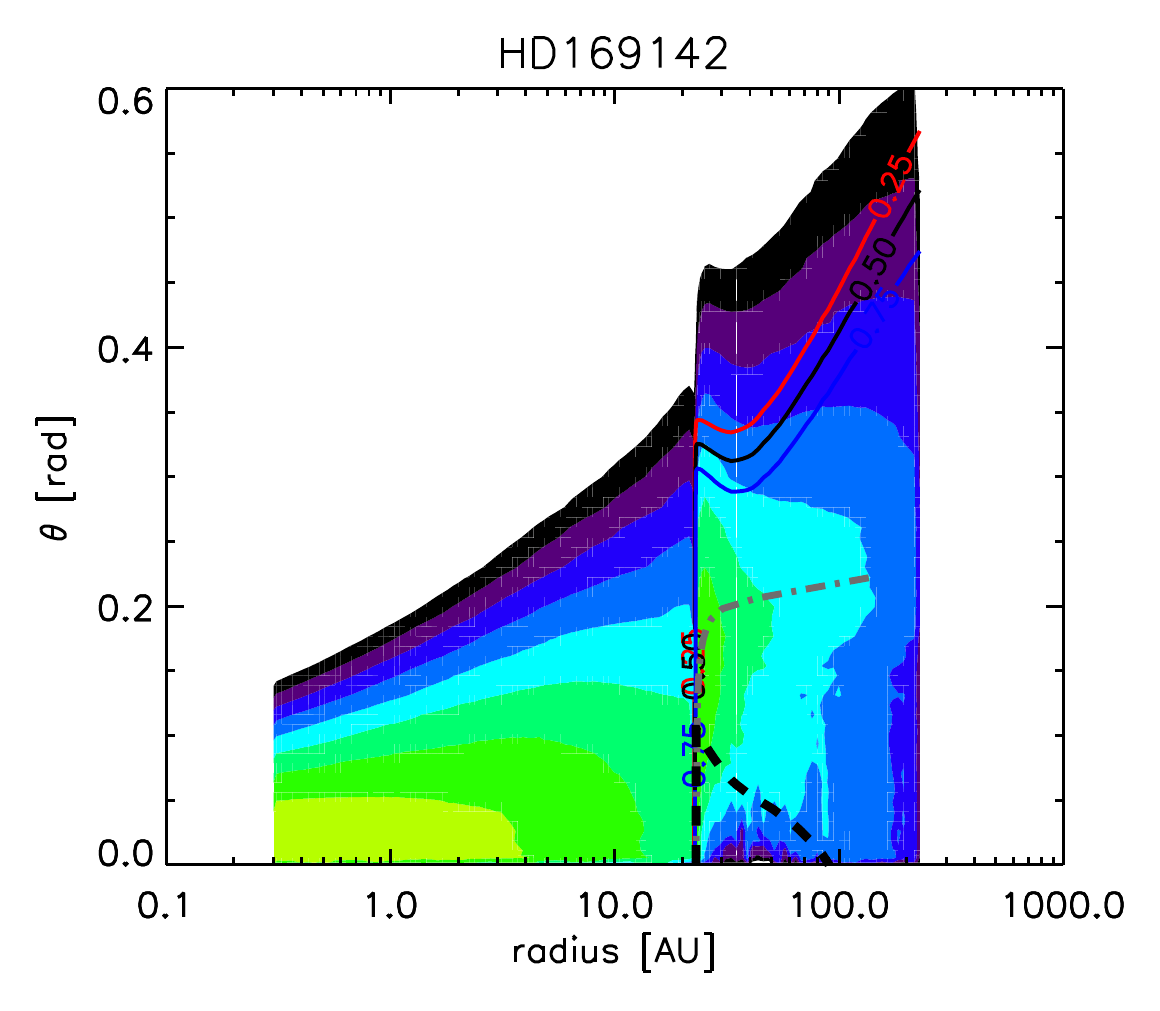}
\includegraphics[width=\columnwidth]{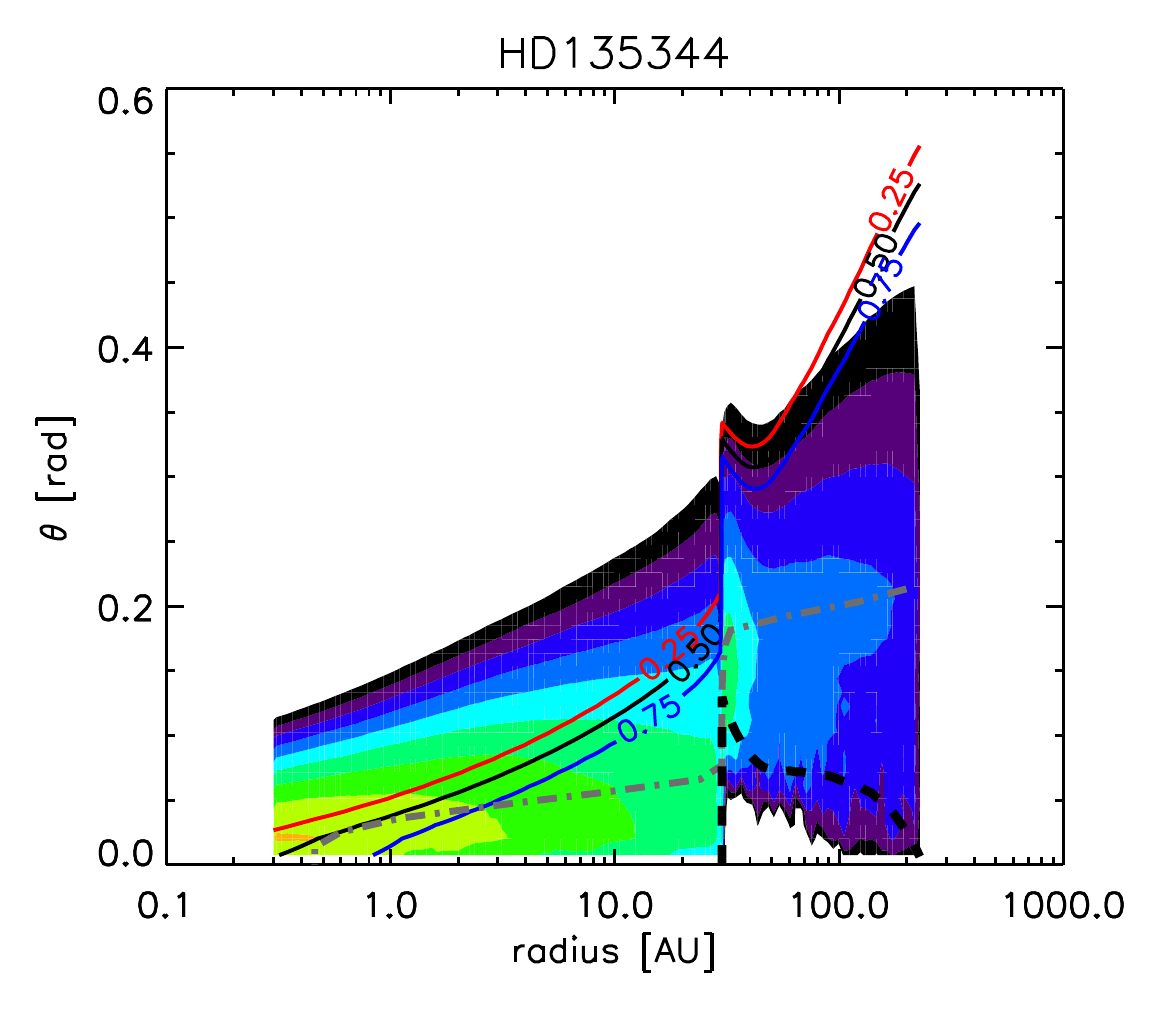}
\includegraphics[width=\columnwidth]{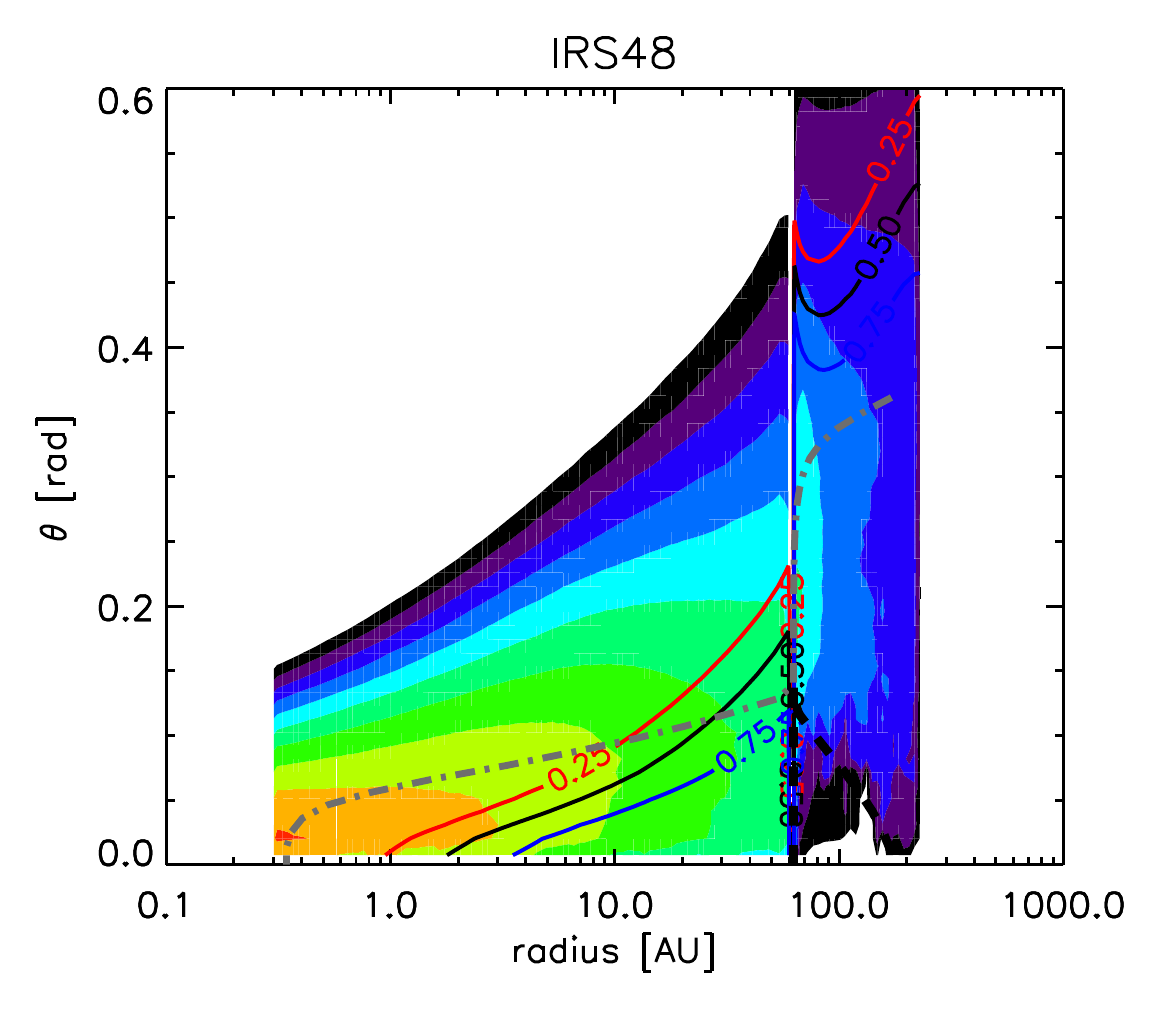}

\caption{ \label{fig:diagnosticplotstransitional}Diagnostic plots showing the origin of the PAH flux in the disk, i.e. the contributions of PAH emission to the final spectrum. The intensity of the PAH flux is shown in logarithmic scale (arbitrary scaling), where each colour spans an order of magnitude. Top left: HD\,97048, top right: HD\,169142, bottom left: HD\,135344\,B, bottom right: Oph IRS 48. The red, black and blue solid lines give the locations where the neutral fractions $f(0)$ are respectively 0.25, 0.5, 0.75. The neutral fractions are computed by equations \ref{eq:f0} and \ref{eq:gamma} in each grid cell of the disk. PAHs in the surface of the disk are largely ionized, while PAHs in the mid-plane are neutral. The black dashed line is the vertical $\tau_{\mathrm{MIR}}=1$ surface at 10 $\upmu$m, the grey dashed-dotted line gives the radial $\tau_{UV}=1$ surface at 0.1 $\upmu$m. The height $\theta$ relates to Z by, $\theta = \tan(Z/r)$. }
\end{figure*}

Our model study in Section \ref{sec:results} has shown that the PAH emission from disks is predominantly neutral in character unless the disk mass is very low. In a more general sense, ionized PAHs are produced by high UV fields (or low dust densities) and low electron densities. The fitting of the observed $I_{6.2}/I_{11.3}$ ratio of PAHs in the integrated spectrum of the disk has been done by adding a low-density gaseous disk in the gap. In this way, PAH emission originates not only in predominantly neutral PAHs in the optically thick inner and outer disk but also in ionized PAHs in the disk gap. To model the gas flows though the gap and to prevent PAH self-shielding effects, we assume a radial dependence of the surface density of $\Sigma \propto r^{0}$ in the gap. The gas surface densities $\Sigma_{gas}$ as a function of radius are shown in Figure \ref{fig:surf_dens}. The required surface densities of the gas within the gap are on the order of $\Sigma_{gas} \sim 10^{-3}-10^{-4}$ g cm$^{-2}$, which are consistent with hydrodynamical predictions of gas flows through planets opening a gap (e.g. \citealt{1999Bryden, 2011Dodson-Robinson, 2012Zhu, 2013Fung}). The basic parameter assumptions and results of our best-fit models are given in Table \ref{tab:fourmodels}. Figure \ref{fig:diagnosticplotstransitional} shows the PAH intensities of our derived PAH models as a function of the location in the disk.

\subsection{PAH fit of the transitional disk HD\,97048 }

The PAH luminosity of HD\,97048 can be fitted with a PAH-to-dust mass fraction of $5.0\times10^{-4}$ by giving a luminosity of $L_{PAH}/L_{*}=9.13\times10^{-3}$, which is close to the observed 9.01 $\pm$ 0.03 $\times$ 10$^{-3}$ in the IRS/Spitzer spectrum. The averaged neutral fraction is $f(0) =0.97$; thus, only 3\% of the PAHs in the disk are ionized. 
%The neutral fraction has increased compared to the benchmark model since PAHs in the wall (i.e. the inner edge of the outer disk) are in a denser environment compared to the surface of a disk. Thus the electron density is higher and PAHs can recombine more easily into the neutral state. 
We note that the calculated $I_{6.2}/I_{11.3}$ ratio (1.27) is larger than the observed one (0.87). However, that may well be an `artefact' of our adopted PAH properties as the intrinsic $I_{6.2}/I_{11.3}$ ratio of neutral PAHs in the \citet{2007DraineLi} model is arbitrarily set.  Figure \ref{fig:parametergrid} shows that the benchmark model in which 100\% of the PAHs are neutral, which still has a higher $I_{6.2}/I_{11.3}$ fraction (0.94) than HD\,97048. Thus, a better fit to the the observed  $I_{6.2}/I_{11.3}$ ratio of HD\,97048 is not possible because the adopted PAH properties do not cover lower $I_{6.2}/I_{11.3}$ ratios.

\begin{figure}[t]
\includegraphics[width=\columnwidth]{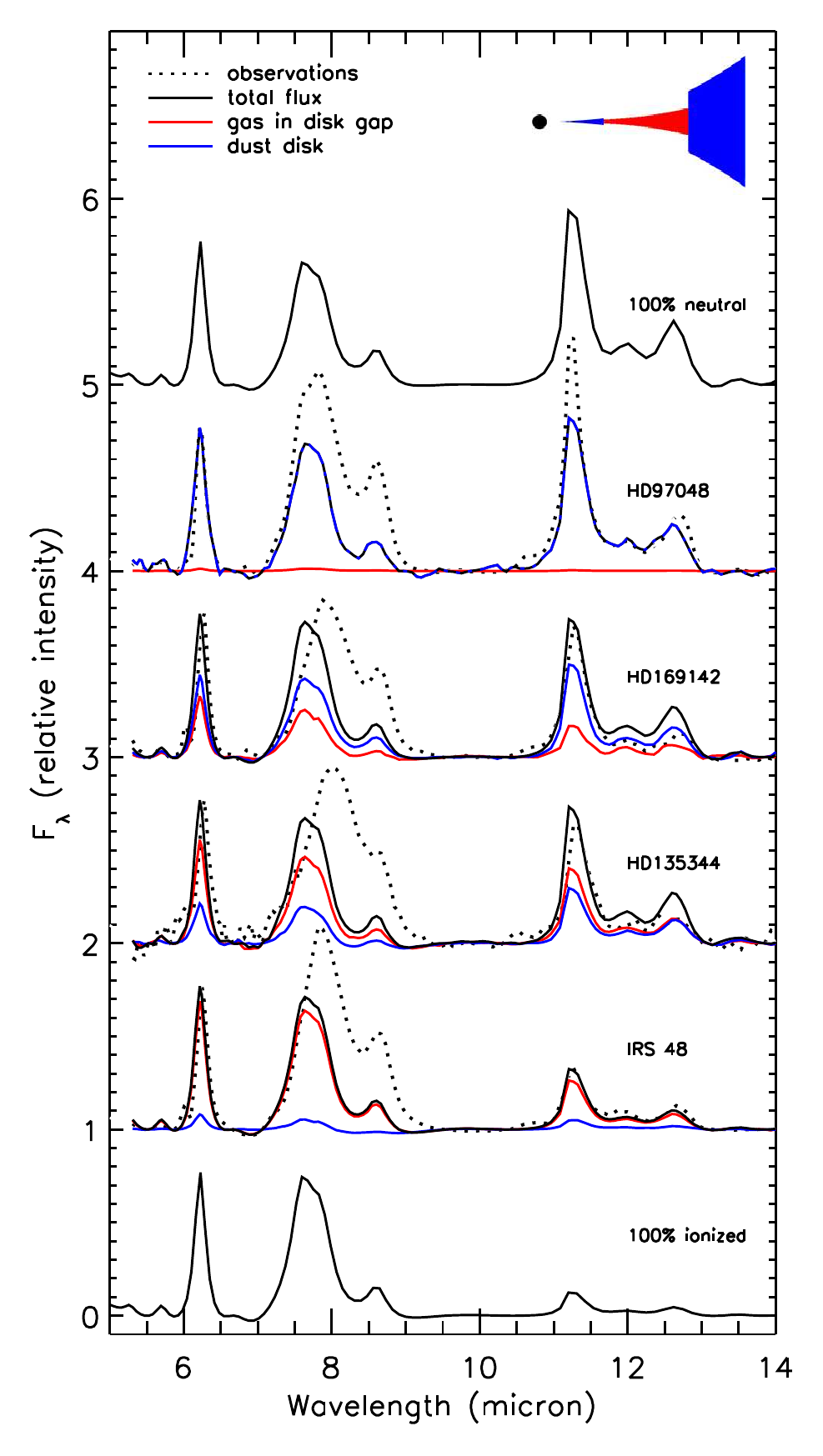}
\caption{ \label{four_objects_PAHs_model} The spectra at the top and the bottom represent the benchmark models with respectively 100\% neutral and 100\% ionized PAHs. The middle four spectra show transitional disks with the dotted black line the Spitzer observations, the solid black lines the total fluxes of the best fit models, the solid blue line the contribution from PAHs in the dust disk, and the solid red line the contribution from the low density optically thin disk. Note that for HD\,97048, there is no low density optically thin disk component  thus the solid black and blue lines are merged into a blue-black dashed line. For Oph IRS 48, the spectrum is dominated by ionized PAHs from the low density optically thin disk.}
\end{figure}

\begin{figure}[t]
\includegraphics[width=\columnwidth]{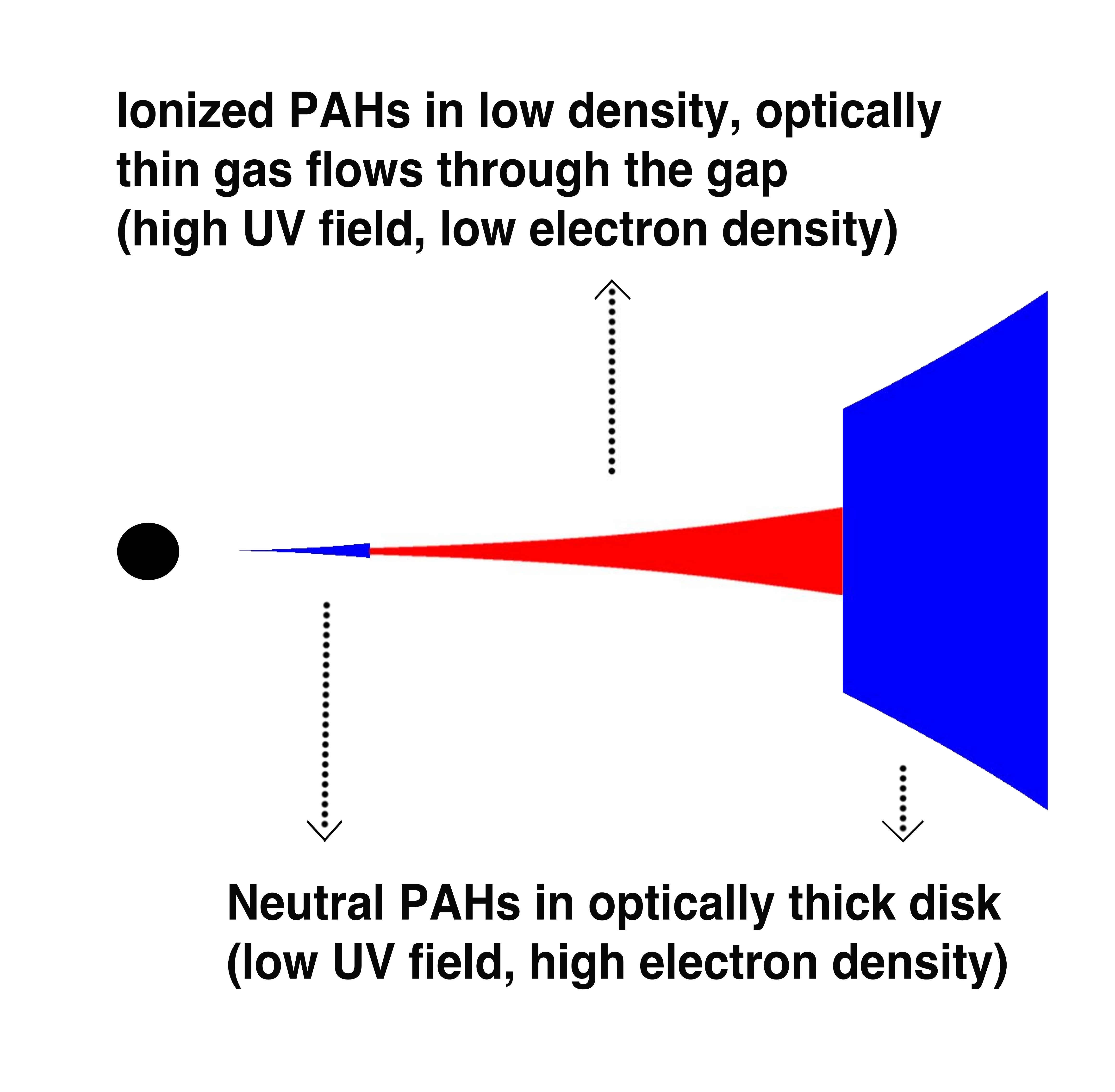}
\caption{ \label{fig:sketch} Sketch of the main result of this paper. Neutral PAH emission originates from an optically thick disk (blue components) and ionized PAH emission from low density, optically thin gas flows through protoplanetary disk gaps (red component).  }
\end{figure}

The gap of HD\,97048 may not be completely empty because no gap is seen in polarized scattered light \citep{2012Quanz}. Therefore, we fill the gap of HD\,97048 with a gas surface density which of $\Sigma_{gas} =  2.5\times10^{-4}$ g cm$^{-2}$, which is similar to the solutions we find for the other transitional disks in our sample (in the next Section \ref{sec:PAHfitstransitionaldisks}). Due to the optically thick inner disk, most of the material in the gap is in the shadow of the inner disk. Thus, the PAHs in the disk gap only contribute 5\% to the total PAH luminosity. A small fraction of dust may also be present in the gap, which may explain the scattered light observations of \citet{2012Quanz}. However, we do not further investigate that scenario since that is not the focus of our paper. The spectra of the PAH components from the optically thick dust disks are shown by the blue lines in Figure \ref{four_objects_PAHs_model}. The red lines show the PAH emission components from the low-density optically thin gas flows in the gaps.

There are two possibilities why HD\,97048 is largely neutral. One explanation is because less (ionized) PAH emission originates in the gap. This occurs because they are in the shadow of the optically thick inner disk. The second explanation is because the contribution from PAHs in the large flaring outer disk is still large and dominating (compared to the other disks in our sample which have smaller outer disk radii). We have run several test models without optically thick inner disks, test models with a higher surface density in the gap, and test models with smaller outer disk radii. The results suggest that the first explanation is most important, because models without an optically thick inner disk show a strong increase of $L_{PAH}$ and $I_{6.2}/I_{11.3}$ due to the fact that (ionized) PAHs in the gap are now directly exposed to the UV emission. Since this is not observed, we suggest that the inner disk of HD\,97048 is still optically thick.

\subsection{PAH fits of the transitional disks HD\,169142, HD\,135344\,B and Oph IRS 48} 
\label{sec:PAHfitstransitionaldisks}
The model solution of HD\,97048 does not work for  HD\,169142, HD\,135344\,B, and Oph IRS 48 for two reasons. First, the model is not consistent with the FWHM profiles because the observed PAH features are not extended compared to the continuum. Second, the observed $I_{6.2}/I_{11.3}$ ratios are too high to be fitted by neutral PAHs from the outer disk suggesting that a significant contribution comes from ionized PAHs. To fit the PAH spectra of HD\,169142, HD\,135344\,B, and Oph IRS 48 we increase the contribution from the low density, optically thin gas in the disk gaps. We keep the PAH-to-dust fraction in the outer disk similar to that of HD\,97048 ($f_{PAH,disk}=5.0\times10^{-4}$).

To obtain a high PAH luminosity and low dust continuum, we assume that the opacities are dominated by the PAHs in the gap. For that reason, we assume that there are no small ($\sim1 ~\upmu$m) grains and, thus, no dust is visible in the spectrum. To get a lower limit to the gas mass needed in the disk gap, we adopt a PAH mass fraction similar to the ISM of $f_{PAH,gap}=4.0\times10^{-2}$. For HD\,169142, HD\,135344\,B and Oph IRS 48, the luminosity $L_{PAH}/L_{*}$  and the $I_{6.2}/I_{11.3}$ ratio are simultaneously fitted by adjusting the total amount of mass in the gas disk in the gap. The characteristics are given in Table \ref{tab:fourmodels}. We have aimed to fit $L_{PAH}/L_{*}$  and $I_{6.2}/I_{11.3}$ as close to the observed values as possible. We find that models could be found, which fit the observed $L_{PAH}/L_{*}$  and $I_{6.2}/I_{11.3}$ within a factor of 2 for all the objects. Following this fitting procedure, PAHs in the gas flows through the gap produce respectively 33\%, 73\%, and 91\% of the total PAH luminosity of HD\,169142, HD\,135344\,B, and Oph IRS 48. 

The diagnostic plots in Figure \ref{fig:diagnosticplotstransitional} show that the disk gap of HD\,169142 has a high fraction of ionized PAHs, while the disk gap of HD\,135344\,B has a much larger fraction of neutral PAHs. This difference can be attributed to the lower stellar temperature of HD\,135344\,B ($T = 6590$ K)  compared to HD\,169142 ($T=8200$ K) producing a lower UV-field. 

\subsection{Summary}

We have found a model solution to fit $L_{PAH}$ and the $I_{6.2}/I_{11.3}$ ratio of the transitional disks HD\,97048,  HD\,169142, HD\,135344\,B, and Oph IRS 48. Observations constrain a contribution of ionized PAHs that arise from within the dust gap (see Figure \ref{four_objects_PAHs}). Our PAH models combine PAH contributions from the surface of a flaring dust disk and a gaseous disk component inside the gap where the ionization fraction can be high. The density structure of gas and dust  in the disk is key to the PAH charge state, since the UV field and electron density are the main parameters that determine the PAH ionization. 

We stress that the scale height and surface densities as a function of radius are poorly constrained for the inner regions. However, the presented PAH model solutions do make a constraint that a particular abundance of ionized PAHs in an optically thin, low density environment is required to fit the spectra. We have chosen to present the models for which the vertical density structure is set by hydrostatic equilibrium. Alternatively we can parameterize the density structure in the disk gap and adopt a higher scale-height of the gas in the gap to obtain similar results. For example, an even better fit to the observed $L_{PAH}/L_{*}$ can be obtained by artificially increasing the scale height of the gas in the gap for Oph IRS 48 (which may be an effect of a larger gas temperature due to UV irradiation). A fit to the observations is then found using a slightly higher mass $M_{PAH,gap} = 1.2\times10^{-9} \ \mathrm{M}_{\odot}$  and setting the vertical scale height $H_0= 3.0$ AU at $R = 30$ AU where $H\propto r^{ 1.1}$. For comparison, the hydrostatic disk structure has a pressure scale height of $H_0 = 2.1$ AU at $R = 30$ AU.

%-----------------------------------------------------------------------------------------
%					DISCUSSION
%-----------------------------------------------------------------------------------------

\section{Discussion}
\label{sec:discussion}

In the following section we discuss our results on PAH ionization in relation to the disk structure. First, we discuss observations and models of gas in protoplanetary disk gaps. Thereafter, we discuss our results in the context of disk evolution in Herbig stars.

\subsection{Gas flows through protoplanetary gaps}

To fit the observed PAH spectra of the transitional disks in our study, we find that a requirement is to add a `gaseous' low-density, optically thin disk in the gap to account for a higher fraction of ionized PAHs. Observations of gas in disk gaps can be understood in the context of planet formation models. The disk should go through a gas-rich/dust-poor stage if planets form by a two-step process where grains first agglomerate into large rocky cores, followed by the accretion of a gaseous envelope (e.g. \citealt{1993Lissauer}).  As PAHs are small molecules, which are thought to travel along with the gas, we first compare PAHs to other gas diagnostics. Thereafter, we discuss our results in the context of theoretical  modeling of gas flows through gaps.

Recent Herschel/PACS observations show that Herbig objects with pure rotational CO transitions have a strong UV luminosity and strong PAH emission \citep{2013Meeus}. However,  for a number of other molecular lines, no connection to the PAH luminosity has been reported \citep{2013Fedele}. From these observations, it is difficult to deduce how PAHs and the gas are connected. On the other hand, there is ample evidence for (molecular) gas in disk gaps. In high-resolution M-band spectral studies of transitional disks, modeling of the rovibrational CO emission yields a great diversity of inner disk gas content and of gas/dust ratios \citep{2007Salyk, 2009Salyk}. Similar studies have been carried out for HD\,135344\,B  \citep{2008Pontoppidan} and Oph IRS 48 \citep{2012aBrown} where the respective authors find evidence for significant quantities of CO gas within the dust gap. \citet{2011Salyk} show that the detection efficiency of H$_2$O, OH, HCN, C$_2$H$_2$, and CO$_2$ depends on the disk color,  $F_{30}/F_{13.5}$, H$\alpha$ equivalent width, and, tentatively, the mass accretion rate, $\dot{M}$. This may suggest a connection between regions of lower dust/gas ratio in the disk and the strength of these molecular lines. Recent ALMA observations of HD\,142527 have shown that reservoirs of gas are present in the dust depleted regions \citep{2013Casassus}. These observations show that gas flows through dust-poor regions are not uncommon in protoplanetary disks, which is consistent with our results. 

Hydrodynamical simulations of the disk structure caused by planet formation predict the opening of an annular `gap', where the structure and depth of the gap is dependent on the mass of the planet and disk viscosity (e.g., \citealt{1999Bryden, 2004Varniere}). Mass can penetrate through the gap and flow onto the planet, or flow through the gap and replenish the inner disk \citep{1999Lubow,2004Paardekooper,2008EdgarQuillen, 2011Tatulli}. Since pressure gradients at the outer edge of the gap, which are cleared by the planet, act as a filter, this process can also produce a very large gas-to-dust ratio in the inner disk, potentially explaining (pre-)transitional disks, which have optically thin inner cavities but still have relatively high accretion rates \citep{2006Rice,2011Mendigutia}. Recent work by \citet{2012Pinilla} demonstrates how dust particles are trapped further out than the location of the edge of the gap in the gas for a given planet. For example, the outer edge for the gas would be located at 37 AU in the specific case of a 15 M$_{\mathrm{Jup}}$ planet at 20 AU, while the dust ring would be at 54 AU. In these models, the surface densities of the gas in the gap can drop to $ \Sigma_{gas} \sim 10^{-3}-10^{-4}$ g cm$^{-2}$, which is similar to other studies on the gas densities in the gaps of transition disk caused by the presence of (multiple) planets (e.g., \citealt{1999Bryden, 2011Dodson-Robinson, 2012Zhu, 2013Fung}). In a paper by \citet{2013Ovelar} using the models of \citet{2012Pinilla}, it is demonstrated how multi-wavelength observations of the gap size (i.e., the location of the inner edge of the outer disk) may help to constrain the planet mass. The results of this paper may further help to constrain the physical properties of the gap. The required surface densities in the gaps of our models are $\Sigma_{gas}\sim10^{-3}-10^{-4}$ g cm$^{-2}$, which are consistent with the surface densities found in hydrodynamical simulations. The ionization balance of PAHs can, therefore, serve as an excellent diagnostic tool to characterize the gas flows through the gaps.

\subsection{PAH ionization and Herbig star disk evolution. }
Our study shows that neutral PAHs originate in optically thick environments where the UV field is low, and the electron density is high. Ionized PAHs originate in low density, optically thin gas flows where the UV field is high and the electron density is low. Figure \ref{fig:sketch} gives a summary sketch of these two disk environments in a transitional disk. 

The two most extreme $I_{6.2}/I_{11.3}$  ratios among Herbig stars come from HD\,141569, where the PAHs are likely almost all ionized, and HD\,97048, where the PAH spectrum represents the highest neutral fraction (Table \ref{table_sed_parameters}). In HD\,141569, studies have shown that the disk is largely optically thin \citep{2003LiLunine,2013Thi}, iwhich agrees with our result that ionized PAHs originate in optically thin environments. In HD\,97048, neutral PAHs originating in the optically thick disk dominate. Even though there may be gas in the gap of HD\,97048 due to the higher density of an optically thick inner disk, the PAHs in the low-density gap are not exposed enough to the UV field.

The effect of the disk geometry on the ionization balance may be visible in the $I_{6.2}/I_{11.3}$ ratio compared to the MIR spectral index $F_{30}/F_{13.5}$ (Figure \ref{fig:ion_vs_LPAH}). Flaring/transitional (group I) objects have a wider range in their degree of ionization compared to flat (group II) objects. This can be interpreted, as that PAH emission in flaring/transitional objects originates in more varied physical environments: those that are predominantly neutral in the optically thick disk and predominantly ionized in optically thin gaseous disk gaps. The PAH emission from flat disks show more similar $I_{6.2}/I_{11.3}$ ratios. This suggests that the disks of flat (group II) objects are more homogeneous. Possibly this could be due to the lack of large gaps in these objects. This would be consistent with the scenario that flaring (group I) objects are often \mbox{(pre-)transitional}, while self-shadowed (group II) objects lack any signatures of large dust gaps.

While ionized PAHs do trace low density gas flows, it is not always the case that gas flows have ionized PAHs. For example, the Herbig stars HD\,135344\,B and HD\,142527 have rather low temperatures (respectively $T= 6590$ K and $T= 6260$ K), and, therefore, their UV fields are less efficient in ionizing the PAHs. Our models show that PAHs will be highly neutral in the disks of T Tauri stars.
 Thus, ionized PAHs are most likely found in low density gaps around Herbig stars with temperatures of $T\gtrsim6000$K.  Alternatively, if the gas density in the gas flow is high, then the PAHs may recombine more easily into the neutral state. This scenario may be part of the explanation why the spectrum of HD\,97048 is still largely neutral. 

A correlation is found between the mm luminosity and the $I_{6.2}/I_{11.3}$ ratio (Figure \ref{fig:6_2_divided_by_11_3_vs_L1300_divide_Lstar1300_divideL}). Since the disk is optically thin at millimeter wavelengths, the mm luminosity may be a proxy of the disk mass in mm-sized dust grains. Therefore, for lower disk masses, the fraction of ionized PAHs seems to be higher. There are two possible ways to understand this trend. The first is that the inner region of the disk becomes optically thinner when accretion depletes most of the inner disk mass and less material from the outer disk is available to replenish the inner regions. As a consequence, the inner disk becomes optically thin, the low density gas flows become more irradiated by UV emission, and the ionization fraction is higher. Another possibility is that more planets are formed creating larger gaps over the course of time when the disk is loosing its mass. Consequently, more low density gas flows emerge with a high fraction of ionized PAHs.

%-----------------------------------------------------------------------------------------
%					CONCLUSIONS
%-----------------------------------------------------------------------------------------

\section{Conclusions}
\label{sec:conclusions}
We have modeled the charge state of PAHs in Herbig disks. We find that the neutral fraction of the disk is high for typical star and disk parameters. The key parameter to ionize the PAHs in the disk is to lower the gas density. The PAH luminosity is most influenced by competition from dust grains, either by an increased PAH mass fraction, or the removal of small dust grains. Our results can be summarised as follows: 

\begin{itemize}
\item Emission from ionized PAHs trace low-density optically thin regions and has a high $I_{6.2}/I_{11.3}$ ratio. 
\item The PAHs in an optically thick disk are predominantly neutral and produce emission with a low $I_{6.2}/I_{11.3}$ ratio.
\item The PAH spectra of transitional disks can be understood as superpositions of neutral PAHs from an optically thick disk and ionized PAHs from that of low density, optically thin gas flows through protoplanetary disk gaps.
\item There is no gradient in the neutral fraction of PAHs as a function of radius in an optically thick disk. 
\item The PAH luminosity is not synonymous to the disk structure. As shown for the transitional disks in our sample, the total PAH mass in the low density, optically thin disk gap can be much smaller compared to the total PAH mass in the outer disk. Nevertheless, they give a dominating contribution to the total PAH luminosity. 
\item We find a trend in the $I_{6.2}/I_{11.3}$ ratio compared to the 1.3 millimeter luminosity. Since the 1.3 millimeter luminosity can be used as a proxy for the disk mass, the degree of ionization tends to be higher for lower disk masses. We suggest that the contribution to the PAH luminosity from ionized PAHs in low-density optically thin `gas flows' becomes larger for lower mass disks. 
\item For a sample of 18 Herbig stars, the $I_{6.2}/I_{11.3}$ ratios indicates that PAH emission from flaring/transitional disks (group I) is higher and originates in a wider range of physical environments  \mbox{($\langle I_{6.2}/I_{11.3} \rangle = 2.61 \pm 1.63$)}, while the origin of PAH emission in self-shadowed flat disks (group II) seems to be more homogeneous \mbox{($\langle I_{6.2}/I_{11.3} \rangle = 2.08 \pm 0.56$)}. 

\end{itemize}

With superior mid-IR sensitivity and spatial resolution, observations by the Mid-InfraRed Instrument (MIRI) on the James Webb Space Telescope (JWST) will be able to probe more and fainter targets. This work shows that the ionization balance of PAHs  provides an extra diagnostic tool to characterize the gaps of transitional disks and, thus, to study the process of planet formation.

\begin{acknowledgements}
The authors thank the referee Ralf Siebenmorgen for useful comments which helped to improve the paper. The authors thank Bram Acke for providing the reduced data from his 2010 paper. The authors thank Carsten Dominik and Alessandra Candian for insightful comments and discussions which improved the analysis presented in this paper. The authors thank Lucas Ellerbroek for proof-reading and providing very useful comments. The authors thank the editor Malcolm Walmsley for providing useful comments on the final version of this paper. K.M. is supported by a grant from the Netherlands Research School for Astronomy (NOVA).  M.M. acknowledges funding from the EU FP7-2011 under Grant Agreement No 284405.  Studies of interstellar chemistry at Leiden Observatory are supported through advanced-ERC grant 246976 from the European Research Council, through a grant by the Dutch Science Agency, NWO, as part of the Dutch Astrochemistry Network, and through the Spinoza premie from the Dutch Science Agency, NWO.
 \end{acknowledgements}

\bibliography{2014-02-04_PAH_ionization_as_a_tracer_of_disk_evolution.bib}

\clearpage

\end{document}